# Double diffusion encoding and applications for biomedical imaging

Rafael N. Henriques[1], Marco Palombo[2], Sune N. Jespersen[3,4], Noam Shemesh[1], Henrik Lundell[5], Andrada Ianuş[1*]

[1] Champalimaud Research, Champalimaud Centre for the Unknown, Lisbon, Portugal
[2] Centre for Medical Image Computing and Dept of Computer Science, University College London, London, UK
[3] Center of Functionally Integrative Neuroscience (CFIN) and MINDLab, Department of Clinical Medicine, Aarhus University, Aarhus, Denmark
[4] Department of Physics and Astronomy, Aarhus University, Aarhus, Denmark
[5] Danish Research Centre for Magnetic Resonance, Centre for Functional and Diagnostic Imaging and Research, Copenhagen University Hospital Hvidovre, Denmark
* corresponding author: Andrada Ianuş, andrada.ianus@neuro.fchampalimaud.org

## Abstract

Diffusion Magnetic Resonance Imaging (dMRI) is one of the most important contemporary non-invasive modalities for probing tissue structure at the microscopic scale. The majority of dMRI techniques employ standard single diffusion encoding (SDE) measurements, covering different sequence parameter ranges depending on the complexity of the method. Although many signal representations and biophysical models have been proposed for SDE data, they are intrinsically limited by a lack of specificity. Advanced dMRI methods have been proposed to provide additional microstructural information beyond what can be inferred from SDE. These enhanced contrasts can play important roles in characterizing biological tissues, for instance upon diseases (e.g. neurodegenerative, cancer, stroke), aging, learning, and development.

In this review we focus on double diffusion encoding (DDE), which stands out among other advanced acquisitions for its versatility, ability to probe more specific diffusion correlations, and feasibility for preclinical and clinical applications. Various DDE methodologies have been employed to probe compartment sizes (Section 3), decouple the effects of microscopic diffusion anisotropy from orientation dispersion (Section 4), probe displacement correlations, study exchange, or suppress fast diffusing compartments (Section 6). DDE measurements can also be used to improve the robustness of biophysical models (Section 5) and study intra-cellular diffusion via magnetic resonance spectroscopy of metabolites (Section 7). This review discusses all these topics as well as important practical aspects related to the implementation and contrast in preclinical and clinical settings (Section 9) and aims to provide the readers a guide for deciding on the right DDE acquisition for their specific application.

# 1. Introduction

Diffusion Magnetic Resonance Imaging (dMRI) has become a powerful methodology for probing microscopic length scales and shapes in living tissues. In dMRI, magnetic field gradients are harnessed to sensitize the acquired signal towards displacement statistics of the MR-responsive molecules residing within the tissue, which carry abundant information on the underlying tissue microstructure.

The vast majority of dMRI techniques rely on the basic diffusion encoding schemes pioneered by Stejskal and Tanner in 1965 (Stejskal and Tanner 1965) for probing displacement of molecules in solution NMR using a pair of diffusion-sensitizing magnetic field gradients as illustrated in Figure 1a. In the context of MRI, we here use the terminology introduced in (Shemesh, Jespersen et al. 2015) and refer to this kind of diffusion encoding sequence as "single diffusion encoding" (SDE), – in the spin echo form, it is also widely known as pulsed field gradient spin echo (PGSE). The main feature of SDE is a single epoch over which diffusion is encoded. Over the years, SDE-based pulse sequences have been thoroughly investigated with respect to signal representations (Jensen, Helpern et al. 2005, Kiselev 2010) and biophysical models (Assaf, Blumenfeld-Katzir et al. 2008, Alexander, Hubbard et al. 2010, Zhang, Schneider et al. 2012, Fieremans, Benitez et al. 2013) – reflecting a richness offered by the relatively large parameter space spanned by SDE sequences, including diffusion time, gradient magnitude, duration and orientation (Alexander, Dyrby et al. 2017, Ghosh, Ianus et al. 2018, Novikov, Fieremans et al. 2019). In turn, SDE can be employed to map numerous parameters of interest depending on the aim and complexity of the dMRI technique. Bulk diffusion tensors (Basser, Mattiello et al. 1994) have been used successfully for mapping orientation of white matter fibres and for highlighting ischemia, among many other applications (Alexander, Lee et al. 2007 ). Techniques which aim to characterize the underlying tissue microstructure beyond the bulk measures provided by diffusion tensor imaging (DTI), usually employ SDE sequences with multiple gradient strengths (Cohen and Assaf 2002, Alexander, Dyrby et al. 2017, Ghosh, Ianus et al. 2018, Novikov, Fieremans et al. 2019) and/or a good angular resolution (Tuch, Reese et al. 2002). To further estimate the restriction sizes in the tissue, many approaches also vary the diffusion time and possibly gradient duration (Woessner 1961, Woessner 1963, Tanner and Liu 1971, Aslund and Topgaard 2009). Nevertheless, one intrinsic limitation of SDE acquisitions involves its weakness in characterizing heterogeneous systems due to the conflation of mesoscopic and microscopic features within the voxel. Thus, such acquisitions can struggle to accurately characterize complex tissue microstructures which include a distribution of pore sizes and orientations or even their averaged properties. To overcome this limitation and allow for robust parameter estimation, SDE techniques aiming to characterize tissue microstructure make various assumptions (e.g. number of compartments, compartment geometry, functional forms of the fibre orientation distribution, permeability and even diffusivity values, etc), which may or may not hold when imaging different tissue types or in the presence of pathology (Lampinen, Szczepankiewicz et al. 2017, Novikov, Veraart et al. 2018, Henriques, Jespersen et al. 2019, Lampinen, Szczepankiewicz et al. 2019).

Various acquisitions which replace the pulsed gradients of the standard SDE sequence with other gradient waveforms have been proposed in the literature, aiming to provide additional information about the underlying microstructural properties. For instance, oscillating gradients can be designed to probe the temporal diffusion spectrum, which offers information on diffusion on short time scales and can inform on, e.g., the pore surface to volume ratio (Mitra, Sen et al. 1992, Schachter, Does et al. 2000, Reynaud, Winters et al. 2016) or diffusion path tortuosities (Parsons Jr., Does et al. 2005). Gradient waveforms with low frequencies can improve the sensitivity to the diameter of elongated pores in the presence of fibre dispersion and/or gradient orientations which are not orthogonal to the fibre direction (Drobnjak, Zhang et al. 2015, Nilsson, Lasic et al. 2017). Diffusion sequences which further vary the gradient orientation within one measurement can disentangle the effects of a distribution of pore sizes and orientations. Acquisitions such as multiple diffusion encoding (MDE), which concatenate two or more gradient pairs, e.g. (Cory, Garroway et al. 1990, Mitra 1995, Cheng and Cory 1999, Shemesh and Cohen 2011, Avram, Ozarslan et al. 2013), or the more recently proposed B-tensor encoding schemes (Eriksson, Lasic et al. 2013, Lasic, Szczepankiewicz et al. 2014, Szczepankiewicz, Lasic et al. 2014, Westin, Knutsson et al. 2016, Topgaard 2017), have been employed to map microscopic diffusion anisotropy without the conflating effects of orientation dispersion at the voxel scale, with applications ranging from material science, biomedical and medical imaging.

A widely studied class of MDE acquisitions is double diffusion encoding (DDE) which combines in one acquisition two diffusion weighting gradient pairs separated by a mixing time, as illustrated in Figure 1b. This diffusion encoding design enables the exploration of spin dynamics – in particular, correlations *between* displacements – beyond what is achievable with standard SDE acquisitions. DDE sequences have been first theoretically introduced in 1990 by Cory et al. (Cory, Garroway et al. 1990), who proposed to measure the local eccentricity of randomly oriented yeast cells using a combination of DDE with parallel and orthogonal gradients. Later on, the early works of Callaghan on this topic (Callaghan and Xia , Callaghan and Manz 1994) employed DDE sequences with different timings to map local velocity fluctuations and exchange. A first comprehensive theoretical analysis of DDE contrast for fully restricted diffusion was described by Mitra in 1995 (Mitra 1995), who showed a signal modulation with the angle between the two gradient pairs, which is proportional to the size of the compartments. These concepts agree with experiments by Cheng and Cory (Cheng and Cory 1999), who estimated the size and shape of prolate yeast cells based on DDE with parallel and orthogonal gradient orientations.

These theoretical and experimental developments led to an increasing interest in DDE acquisitions, especially for biomedical imaging, to provide a more comprehensive characterization of the underlying tissue microstructure, which is of great importance for studying brain development, plasticity, aging, as well as for the detection and monitoring of various neurological conditions. Thus, over the last two decades many studies have analysed this class of diffusion acquisitions in terms of theory, simulations, preclinical imaging and clinical translation, and some of the early studies have been included in previous reviews (Shemesh, Ozarslan et al. 2010, Finsterbusch 2011, Shemesh, Jespersen et al. 2015, Novikov, Fieremans et al. 2019), which we highly recommend as complementary reading to this work. Nevertheless, since the last comprehensive reviews focusing on both DDE theory and applications almost a decade ago, many new and exciting developments took place.

Here we set out to give an up to date review of DDE with a special focus on biomedical imaging, generally following the different purposes and regimes of DDE acquisitions. First we present an overview of SDE and DDE sequence parameters and signal description (Section 2), then we discuss various DDE applications: imaging pore size and size distributions (Section 3), mapping microscopic anisotropy (Section 4); implications for biophysical modelling (Section 5); mapping diffusion correlation and exchange (Section 6); studying metabolites with MR spectroscopy (Section 7). In the end, we review practicalities related to sequences implementation (Section 8) and discuss future perspective on this topic (Section 9).

# 2. Sequence description and comparison with SDE

This section describes the tuneable parameters of the SDE and DDE gradient waveforms and provides the general signal expressions as a basis for the following Sections which present more specific acquisition regimes and applications.

## 2.1 Single Diffusion Encoding

As illustrated in Figure 1a, SDE sequences are characterized by the gradient strength (G), duration (δ) and orientation ($\hat{\boldsymbol{n}}$), as well as the interval between the onset of the gradient pulses, usually referred to as diffusion time (Δ). For a particle, which moves from position $\mathbf{r_0}$ to position $\mathbf{r_1}$ during the time interval Δ, the phase incurred by an ideal SDE sequence with a short gradient duration δ, the so-called short gradient pulse (SGP) approximation, is $\phi = \gamma\delta\mathbf{G}(\mathbf{r}_0 - \mathbf{r}_1)$, where γ is the gyromagnetic ratio. Thus, the normalized diffusion signal, which is computed as the ensemble average of the acquired spin phases $S = \langle \exp(-i\phi) \rangle = \langle \exp(i\gamma\delta\mathbf{G}(\mathbf{r}_1 - \mathbf{r}_0)) \rangle$, can be expressed as (Callaghan 1991):

$$S = \iint \rho(\mathbf{r}_0) P(\mathbf{r}_1 \,|\, \mathbf{r}_0, \Delta) \exp(i\gamma\delta\mathbf{G}(\mathbf{r}_1 - \mathbf{r}_0)) d\mathbf{r}_0 d\mathbf{r}_1 \tag{1}$$

where $\rho(\mathbf{r_0})$ is the initial spin distribution and $P(\mathbf{r_1}|\mathbf{r_0},\Delta)$ describes the probability that a spin moves from position $\mathbf{r_0}$ to position $\mathbf{r_1}$ in the time interval Δ, and is also known as the diffusion propagator (Callaghan 1991). Eq. (1) is usually rewritten in terms of the diffusion wave vector $\mathbf{q} = \gamma\delta\mathbf{G}$ and the displacement $\mathbf{R} = \mathbf{r_1} - \mathbf{r_0}$, which results in the well-known Fourier relationship between displacement and q-space:

$$S = \iint \rho(\mathbf{r}_0) P(\mathbf{r}_0 + \mathbf{R} \,|\, \mathbf{r}_0, \Delta) d\mathbf{r}_0 \exp(i\mathbf{qR}) d\mathbf{R} = \int \overline{P}(\mathbf{R}, \Delta) \exp(i\mathbf{qR}) d\mathbf{R}, \tag{2}$$

where $\overline{P}(\mathbf{R}, \Delta) \equiv \langle P(\mathbf{r}_0 + \mathbf{R} \,|\, \mathbf{r}_0, \Delta) \rangle = \int \rho(\mathbf{r}_0) P(\mathbf{r}_0 + \mathbf{R} \,|\, \mathbf{r}_0, \Delta) d\mathbf{r}_0$ is the averaged diffusion propagator. From this relationship, we can also appreciate the loss of information in SDE acquisitions due to the averaging of all the spins within the voxel.

### 2.1.1 SDE signal for free diffusion

In the case of free diffusion, the propagator is a Gaussian function with a time dependent standard deviation and has the following expression:

$$P(\mathbf{r}_1 \,|\, \mathbf{r}_0, \Delta) = \frac{\exp\left(-(\mathbf{r}_1 - \mathbf{r}_0)^2 / 4D\Delta\right)}{(4\pi D\Delta)^{3/2}} \tag{3}$$

where D is the diffusion coefficient. Inserting this specific form of the diffusion propagator into Eq. (1) for the SDE with short gradient pulses (Price 2009) yields a mono-exponential relationship $S = \exp(\;\; bD)$, where $b$ is usually referred to as the b-value ($b=\gamma^2\delta^2G^2\Delta$, for SDE in the short pulse approximation). For a general diffusion sequence described by a 3 dimensional gradient waveform **G**(t) applied for a total duration T, the b-value is given by the trace of the b-tensor (Basser, Mattiello et al. 1994), i.e. $b = \sum_{i=1}^{3} b_{ii}$, which is calculated as follows:

$$b_{ij} = \gamma^2 \int_0^T \left[ \int_0^t G_i(t')dt' \int_0^t G_j(t')dt' \right] dt \tag{4}$$

For realistic SDE sequences with finite pulse duration δ, Eq. (4) yields the well-known expression $b = \gamma^2\delta^2G^2(\Delta - \delta/3)$.

### 2.1.2 SDE signal for restricted diffusion

The propagator $P(\mathbf{r}_1 | \mathbf{r}_0, \Delta)$ also has known analytical expressions for diffusion restricted inside single pores of special geometries (parallel planes, cylinders, spheres (Neuman 1974, Balinov, Jonsson et al. 1993) and triangles (Laun, Kuder et al. 2012)). Thus, the restricted diffusion signal for ideal SDE sequences with short pulses (SGP approximation) can be directly calculated following Eq. (1). For SDE sequences with arbitrary gradient duration, the restricted signal for relatively low q values can be computed employing a Taylor or cumulant expansion of Eq. (1), which up to 2nd order in q is commonly referred to as the Gaussian Phase Distribution (GPD) approximation (Neuman 1974, van Gelderen, DesPres et al. 1994, Kiselev 2017). For arbitrary gradient waveforms with low amplitudes, the alternative spectral domain analysis introduced by Stepisnik could be used instead (Stepisnik 1993). For higher q values and/or more general gradient waveforms, the restricted SDE signal can be computed using semi-analytical matrix approaches, for example the matrix method (MM) (Codd and Callaghan 1999, Drobnjak, Zhang et al. 2011, Ianus, Alexander et al. 2016), or the multiple correlation function (MCF) formalism (Grebenkov 2007), which discretize the gradient waveform in piece-wise constant parts. Other recent theoretical approaches model the problem of restricted diffusion from the perspective of particles diffusing under potentials (Yolcu, Memiç et al. 2016, Özarslan, Yolcu et al. 2017). Such signal expressions have been widely used in dMRI techniques which aim to estimate restriction sizes. For more general pore shapes, the restricted SDE signal can be calculated using computational approaches, for example numerical solutions of the diffusion equation (Grebenkov and Nguyen 2013, Novikov, Kiselev et al. 2018) or Monte Carlo simulations (Hall and Alexander 2009, Fieremans, Novikov et al. 2010, Palombo, Ligneul et al. 2016, Palombo, Ligneul et al. 2018, Ginsburger, Matuschke et al. 2019, Palombo, Alexander et al. 2019, Callaghan, Alexander et al. 2020).

## 2.2 Double Diffusion Encoding

As illustrated in Figure 1b, DDE gradient waveforms concatenate two pairs of diffusion sensitizing gradients characterized by unique gradient strengths ($G_1$ and $G_2$), durations ($\delta_1$ and $\delta_2$), orientations ($\hat{\boldsymbol{n}}_1$ and $\hat{\boldsymbol{n}}_2$), diffusion times ($\Delta_1$ and $\Delta_2$) and a mixing time ($\tau_m$) separating the two diffusion encoding intervals. For an ideal DDE sequence with short pulse durations ($\delta \rightarrow 0$ ), Eq. (1) becomes:

$$S(\mathbf{q}_1, \mathbf{q}_2, \Delta, \tau_m) = \iiiint \rho(\mathbf{r}_0) e^{-i\mathbf{q}_1\mathbf{r}_0} d\mathbf{r}_0 \cdot P(\mathbf{r}_1 | \mathbf{r}_0, \Delta_1) e^{i\mathbf{q}_1\mathbf{r}_1} d\mathbf{r}_1 \cdot \\ \cdot P(\mathbf{r}_2 | \mathbf{r}_1, \tau_m) e^{-i\mathbf{q}_2\mathbf{r}_2} d\mathbf{r}_2 \cdot P(\mathbf{r}_3 | \mathbf{r}_2, \Delta_2) e^{i\mathbf{q}_2\mathbf{r}_3} d\mathbf{r}_3 \quad (5)$$

where $\mathbf{r}_i$ represents the spin position when the gradients are applied, and $P(\mathbf{r}'|\mathbf{r},t)$ is the probability that a spin moved from position **r** to position **r'** during time interval t. Rewriting Eq. (5) in terms of spin displacements over the two diffusion encoding periods

$$S(\mathbf{q}_1, \mathbf{q}_2, \Delta, \tau_m) = \iiiint \rho(\mathbf{r}_0) P(\mathbf{r}_1 | \mathbf{r}_0, \Delta_1) P(\mathbf{r}_2 | \mathbf{r}_1, \tau_m) P(\mathbf{r}_3 | \mathbf{r}_2, \Delta_2) \\ e^{i\mathbf{q}_1(\mathbf{r}_1 - \mathbf{r}_0)} e^{i\mathbf{q}_2(\mathbf{r}_3 - \mathbf{r}_2)} d\mathbf{r}_0 d\mathbf{r}_1 d\mathbf{r}_2 d\mathbf{r}_3, \quad (6)$$

we see that the DDE signal depends on the convolution of the diffusion propagators $P(\mathbf{r}_1 | \mathbf{r}_0, \Delta_1) P(\mathbf{r}_2 | \mathbf{r}_1, \tau_m) P(\mathbf{r}_3 | \mathbf{r}_2, \Delta_2)$ which is calculated before performing the voxel average,

and thus can provide additional information content compared to the SDE signal, which depends on the voxel-averaged propagator.

### 2.2.1 DDE signal for free diffusion

For free diffusion, the displacement probabilities in Eq. (5) are independent and, as for the SDE signal (c.f. 2.1.1), the DDE signal can be fully described by a mono-exponential as the well-known expression $S = e^{-Db_{DDE}}$, where the b-value $b_{DDE}$ is calculated according to the general expression in Eq. (4). For DDE sequences with $\tau_m > \delta$ (i.e. the 2nd and 3rd pulse do not overlap), the signal becomes the product of the signal decay from the two gradient pairs and $b_{DDE} = b_1 + b_2$, where $b_{1,2}$ are the b-values of the 1st and 2nd gradient pair, respectively. For DDE with zero mixing time (i.e. the 2nd and 3rd pulses overlap), the total b-value will depend on the relative orientation of the two gradient pulses (Finsterbusch 2011). Thus, for a sequence with equal gradient durations $\delta_1 = \delta_2 = \delta$, the b-value is given by $b_{DDE} = b_1 + b_2 + \frac{1}{3}\gamma^2\delta^3 G_1 G_2 \cos\varphi$, where $\varphi$ is the relative angle between the gradient orientations. *Note: the sign in front of cos(φ) is the opposite from the expression in (Ozarslan and Basser 2008), as here we follow the most recent nomenclature , where parallel gradients (i.e. φ = 0) are assumed to have* $\mathbf{q}_2 = \mathbf{q}_1$*, rather than* $\mathbf{q}_2 = -\mathbf{q}_1$ *as previously assumed.* This angular dependence also affects signal measured from free diffusion and should not be confused with the angular dependence due to restricted diffusion, described in the following sections.

### 2.2.2 DDE signal for restricted diffusion

Similar to SDE, the DDE signal can be calculated for diffusion restricted inside pores of special geometries by solving the integral in Eq. (5) for a known propagator $P(\mathbf{r}_1 | \mathbf{r}_0, t)$. If the DDE sequences have arbitrary timing parameters, then analytical expressions of restricted diffusion signal for relatively low q values can be derived using Taylor or cumulant expansions, and results have been presented in (Ozarslan and Basser 2008, Ozarslan 2009) and (Ianus, Drobnjak et al. 2016), respectively. For DDE sequences with higher q values, semi-analytical matrix approaches (Codd and Callaghan 1999, Grebenkov 2007, Ozarslan, Shemesh et al. 2009, Drobnjak, Zhang et al. 2011) are usually employed, while for general pore shapes the DDE signal is usually calculated using numerical methods.

Other theoretical approaches to calculate and understand the DDE signal in different microstructural scenarios have been based on higher order tensors to describe the correlations between the diffusion weighting periods. The work of Finsterbusch and Koch (Finsterbusch and Koch 2008) expresses the second-order Taylor expansion of the signal in terms of a rank-2 tensor, while the more general theoretical approach presented by Jespersen et al (Jespersen and Buhl 2011, Jespersen 2012, Jespersen, Lundell et al. 2013) describes the 4th order cumulant expansion in terms of 4-th order tensors. As demonstrated by Jensen et al. (Jensen, Hui et al. 2014), these latter 4-th order tensors can be parametrized by a single 6th order tensor.

These techniques and their application will be explored in detail in the next sections.

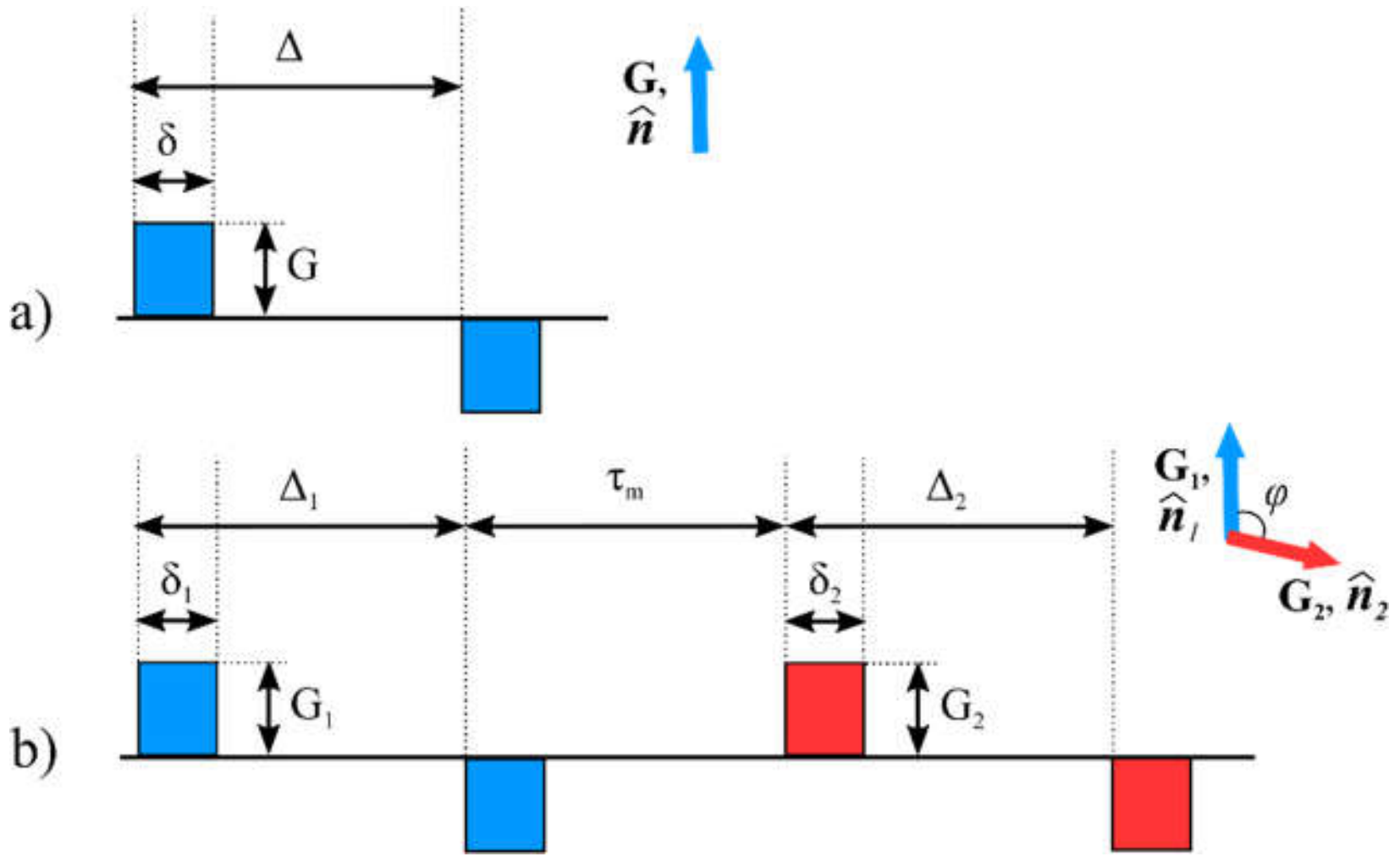

*Figure 1 Effective diffusion gradient waveform and sequence parameters for a) SDE and b) DDE sequences. Depending on the timing parameters and experimental regimes of interest, the diffusion gradients can be added to different spin preparation, such as spin-echo (SE), double spin-echo (DSE), stimulated echo (STE), etc. These aspects will be discussed in the practicalities section.*

# 3. Estimation of pore size

In a historical context, prior to the introduction of DDE, NMR and MRI methods based on SDE with multiple q values (and diffusion times) were the main tools for estimation of compartment sizes in tissues (Cory and Garroway 1990, Stanisz, Szafer et al. 1997, Assaf and Cohen 1999, Assaf, Blumenfeld-Katzir et al. 2008, Ong and Wehrli 2010). Mapping pore sizes is generally interesting in biomedical applications because changes in biological tissues' length scales can inform about brain development, plasticity, as well as various pathological processes which affect the tissue microstructure in disease. For instance, the size (and shape) of normal red blood cells are tightly bound, but in sickle cell anaemia, both the size and shape vary considerably (Pauling, Itano et al. 1949); in patients with multiple sclerosis, small fibres are particularly affected while the large axons are relatively preserved (DeLuca, Ebers et al. 2004); axon diameters (along with myelin content) determine the conduction velocity (Waxman 190); the size of astrocyte bodies can change with activation (Anderson and Swanson 2000) or in the presence of oedema (Kimelberg 1995), etc. However, SDE based methods, as employed in the early q-space imaging (QSI) studies, necessitated very high gradient strength, typically unavailable on clinical systems.

This section describes the main techniques which employ DDE experiments to characterize pore size, why they were considered useful compared with other methods, for example QSI, and how they compare to SDE approaches based on more recent studies. Pore size mapping using DDE has been extensively investigated using theoretical and numerical simulations (Mitra 1995, Ozarslan and Basser 2008, Koch and Finsterbusch 2009, Finsterbusch 2011, Benjamini, Katz et al. 2012, Ianus, Drobnjak et al. 2016) and have been applied to characterize pore sizes in a range of materials from phantoms (Koch and Finsterbusch 2008, Shemesh, Ozarslan et al. 2010, Komlosh, Ozarslan et al. 2011, Shemesh, Ozarslan et al. 2012), ex-vivo (Koch and Finsterbusch 2008, Weber, Ziener et al. 2009, Shemesh and Cohen 2011, Morozov, Bar et al. 2015, Benjamini, Komlosh et al. 2016) and in-vivo tissue (Koch and Finsterbusch 2008, Duchêne, Abarca-Quinones et al. 2020), both using preclinical and clinical scanners.

## 3.1 Low q-value regime: DDE sequences with short mixing time

One of the first comprehensive theoretical descriptions of employing DDE sequences to estimate pore size has been presented by Mitra. For restricted diffusion inside a closed pore, in the limit of long diffusion time, the displacement probability becomes equal to the spins' probability density function, also referred to as the pore space function, i.e. $P(\mathbf{r}'|\mathbf{r},t\rightarrow\infty)=\rho(\mathbf{r}')$. Then, the signal expression for ideal DDE sequences presented in Eq. (5) can be written in terms of the reciprocal pore space function, which is defined as the Fourier transform $\tilde{\rho}(\mathbf{q})=\int d\mathbf{r}\rho(\mathbf{r})\exp(-i\mathbf{q}\mathbf{r})$. Specifically, considering a single pore and a DDE sequence with vanishing mixing time ($\tau_m \rightarrow 0$), for which the positions $\mathbf{r}_1$ and $\mathbf{r}_2$ are the same, Eq. (5) becomes

$$S=\tilde{\rho}(\mathbf{q}_1)\tilde{\rho}(\mathbf{q}_2)^*\tilde{\rho}(\mathbf{q}_2-\mathbf{q}_1), \tag{7}$$

where $\tilde{\rho}(\mathbf{q})^*=\tilde{\rho}(-\mathbf{q})$. For an ensemble of pores, the signal can be calculated by summing the individual contributions $\sum_n \tilde{\rho}_n(\mathbf{q}_1)\tilde{\rho}_n(\mathbf{q}_2)^*\tilde{\rho}_n(\mathbf{q}_1-\mathbf{q}_2)$. Expanding the signal expression in Eq. (7) for small values of $\mathbf{q}$ reads as

$$S(\mathbf{q}_1,\mathbf{q}_2,\Delta\rightarrow\infty,\ \tau_m\rightarrow 0)\approx 1-\frac{1}{3}q^2\left\langle R_{gyr}{}^2\right\rangle(2-\cos\varphi), \tag{8}$$

where φ is the relative angle between the gradient orientations, and $\left\langle R_{gyr}{}^2\right\rangle$ is the mean radius of gyration of the pore $\left\langle R_{gyr}{}^2\right\rangle=\int r^2 d\mathbf{r}$. *Note: the sign in front of cos(φ) is the opposite from (Mitra 1995, Ozarslan 2009) as here we follow the most recent sign convention from (Shemesh, Jespersen et al. 2015).* Explicit signal equations for different restricting geometries have been provided in (Cheng and Cory 1999, Ozarslan 2009, Ianus, Drobnjak et al. 2016). To further account for the macroscopic anisotropy of the system, Finsterbusch and Koch proposed a tensor formalism to replace the scalars in Eq. (8) (Finsterbusch and Koch 2008), and in a subsequent study Finsterbusch showed that concatenating multiple gradient pairs can further increase the amplitude modulation (Finsterbusch 2009).

An example of the angular signal dependence for DDE with short mixing time is illustrated in Figure 2a for three different microstructural substrates with the same radius of gyration, which indeed show a very similar cosine modulation of the signal. The simulated signal has been calculated based on the extension of the matrix approach (Codd and Callaghan 1999) for 3D gradient waveforms (Drobnjak, Zhang et al. 2011) implemented in the MISST package (https://www.nitrc.org/projects/misst/) (Ianus, Alexander et al. 2016). Showing that DDE acquisitions with short mixing times are sensitive to pore sizes even for low q-values, this framework prompted an increasing interest to apply such acquisitions for characterizing restriction sizes.

Although the theoretical equations presented above provide an intuitive description of the DDE signal behaviour when diffusion is restricted, they hold only for ideal sequences with short gradient duration and long diffusion time. In practice, typical DDE acquisitions do not satisfy these assumptions, and employing these equations directly might lead to biased parameter estimates, even using preclinical scanners where experimental conditions are closer to the ideal settings (Shemesh, Ozarslan et al. 2009). Thus, the majority of studies must account for the finite gradient durations, mixing times, and diffusion times using one of the signal calculation methods described in section 2.2.2. Then, the inverse problem can be solved to estimate the pore size from the measured data.

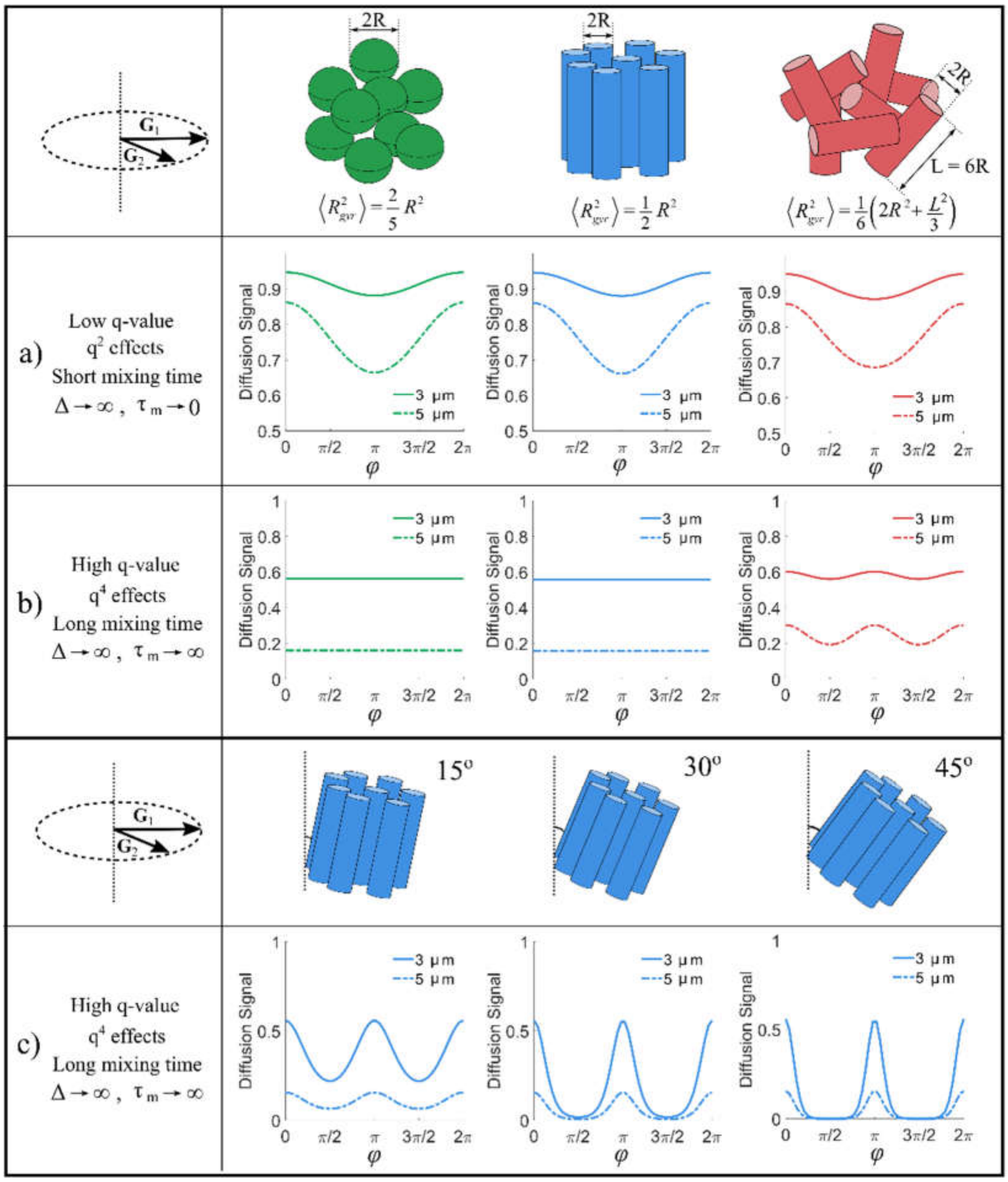

*Figure 2 Simulated angular signal dependence of DDE sequences with a) short mixing time and b) long mixing time for substrates consisting of spheres, parallel cylinders and randomly oriented cylinders with the same radius of gyration of 3 μm (continuous line) and 5 μm (dotted line) with intrinsic diffusivity D = 2 μm²/ms. The DDE parameter values used in the simulation are the following: a) δ = 1 ms, Δ = 100 ms, $\tau_m$ = 1 ms, G = 0.4 T/m and b) δ = 1 ms, Δ = 100 ms, $\tau_m$ = 100 ms, G = 1 T/m. A much stronger gradient strength is used for b, in order to probe the angular dependence of the $q^4$ term. c) Simulated angular signal dependence of DDE sequences with long mixing time for macroscopically anisotropic substrates of parallel cylinders tilted at different angles with respect to the plane of diffusion gradients. The sequence parameters are the same as in b).The simulations have been performed with the MISST software.*

### 3.1.1 Phantom validation

Validation of DDE signals is imperative, especially in the context of biomedical imaging. While Cheng and Cory validated DDE on yeast cells (Cheng and Cory 1999), the first experimental study on a clinical system was done by Koch and Finsterbush (Koch and Finsterbusch 2008), who validated angular DDE's ability to measure pore sizes using radish, packed beads, and even spinal cord specimens. Using a preclinical scanner, Shemesh et al (Shemesh, Ozarslan et al. 2009) later employed angular DDE sequences in NMR experiments to estimate pore size in water-filled glass microcapillary arrays with various nominal diameters between 5 to 20 μm and investigated the effects of different experimental sequence parameters. The results showed a good correspondence to the ground truth, especially when the timing parameters of the sequences have been accounted for (Ozarslan and Basser 2008). Moreover, for phantoms with significant susceptibility-driven gradients, SDE and DDE sequences with bi-polar gradients can be used instead to avoid artifacts (Shemesh and Cohen 2011, Morozov, Bar et al. 2013). Subsequent studies (Komlosh, Ozarslan et al. 2011, Ozarslan, Komlosh et al. 2011) showed the feasibility of applying angular DDE sequences with short mixing time to provide voxel-wise estimates of pore diameter for imaging, both in a similar microcapillary phantom and in plant tissue. Shemesh et al (Shemesh, Ozarslan et al. 2009) further investigated the DDE signal behaviour in a phantom containing two diffusing compartments (free isotropic and restricted anisotropic), and showed that accurate compartment size can be estimated via angular DDE only when the free diffusion signal is effectively suppressed (e.g., at stronger diffusion weighting). A good estimation of capillary size was reported both from SDE and DDE acquisitions when the diffusion time was sufficiently long to probe the characteristic length scale of the pores.

Although the majority of phantom experiments have been performed on preclinical scanners, physical phantoms also play an important role for testing new sequences and contrasts on clinical scanners. For instance, the first experimental investigation of mapping pore size based on angular DDE, was performed by Koch and Finsterbusch on a clinical scanner (Koch and Finsterbusch 2008). To account for possible confounding factors, they have tested the effects of various imaging parameters (slice thickness, orientation, and repetition time), on different samples, including two phantoms. The results showed a robust contrast and a good estimation of restriction size in a phantom consisting of glass beads immersed in water, although in a radish phantom with much larger pore sizes (>50 μm) the values were underestimated, reflecting the importance of matching the sequence parameters to the sample characteristics.

The physical phantoms used so far verified that the DDE signal has the expected characteristics in known geometries, nevertheless, they are quite simple and do not mimic the entire complexity of tissue micro-architecture. The potential of DDE approaches to properly characterize the brain microstructure, as well as the interpretation of the estimated metrics, is still an open question and an active area of research.

#### 3.1.2 Preclinical and clinical applications

For biomedical imaging applications, DDE acquisitions have been employed to characterize restriction sizes in a range of tissue types, both on preclinical high field scanners, as well as on clinical scanners.

On a preclinical scanner, to characterize white matter microstructure, Weber et al (Weber, Ziener et al. 2009) imaged ex-vivo rat spinal cord at ultra-high field (17.6T) and estimated apparent cell radii from angular DDE with short mixing time by fitting Eq. (8) Moreover, they also investigated the dependence of estimated radii on the diffusion times. Shemesh and Cohen (Shemesh and Cohen 2011) employed angular DDE experiments to investigate the diffusion properties in both gray matter (GM) and white matter (WM) using ex-vivo porcine optic nerve and brain tissue, although they did not explicitly estimate the restriction size. A subsequent study of the pig optic nerve, which employed a similar acquisition and a geometrical model of the underlying tissue, provided estimates of the apparent averaged axon diameter relatively close to histological values (Morozov, Bar et al. 2015).

A similar angular dependence has been measured in porcine spinal cord on a clinical scanner (Koch and Finsterbusch 2008), although the amplitude modulation was smaller compared to expected theoretical values, likely due to violations of the modelling assumptions and contributions of free diffusion. To account for these effects when estimating restriction size in different ROIs, the signal expression included an additional free diffusion compartment. To account for unknown fibre orientation when estimating the restriction size, Komlosh et al employed a DDE protocol ($\tau_m$=0) with 3 q-values and parallel and anti-parallel gradients with directions uniformly distributed over the hemisphere (Komlosh, Benjamini et al. 2018). Based on this acquisition, they investigated the changes in apparent mean axon diameter in the optical tract for an ex-vivo mouse brain following traumatic brain injury, finding significantly higher values compared to the control, consistent with histological characterization.

A DDE protocol with parallel and anti-parallel gradients has also been applied on a clinical scanner to characterize the white matter microstructure in healthy volunteers (Koch and Finsterbusch 2011). This study by Koch and Finsterbusch showed the feasibility of measuring signal differences in line with theoretical predictions, and estimated the mean squared radius of gyration in the corticospinal tract following Eq. (8), as illustrated in Figure 3b. A subsequent analysis which assumed a cylindrical pore shape and accounted for the finite gradient timing parameters yielded a volume-weighted mean pore diameter of 13 μm. Although larger than histological values (Innocenti, Caminiti et al. 2015), these results were comparable to the ones reported in the literature from other contemporary SDE techniques applied on clinical scanners (Alexander, Hubbard et al. 2010, Horowitz, Barazany et al. 2015, Suzuki, Hori et al. 2016).

DDE acquisitions have been implemented on clinical scanners by several research groups, nevertheless, most clinical applications have focused on estimating microscopic anisotropy (vide infra) rather than pore size and will be discussed in Section 4.

#### 3.1.3 SDE or DDE for pore size estimation?

As described above, DDE acquisitions with short mixing times have been employed in various studies to characterize restricted diffusion and estimate the mean pore size in a wide range of samples and tissue types. Nevertheless, an in-depth theoretical analysis by Jespersen showed that in the typical limit $q^2\langle r^2\rangle \ll 1$ of weak gradient strengths, the same O($q^2$) information about restriction size can be obtained by measuring SDE signals at multiple diffusion times (Jespersen 2012, Jespersen 2013). That is, unlike QSI, which harnesses the q-value dependence of the signal at a given diffusion time for size estimation, SDE at multiple diffusion times can provide insight into the pore sizes from the time dependence perspective at a given q-value. For an ideal DDE sequence with $\Delta_1 = \Delta_2 = \Delta$, the diffusion weighted signal, written in terms of the ensemble average of spin phases reads as:

$$S(\mathbf{q}_1,\mathbf{q}_2) = \left\langle \exp(-i\phi)\right\rangle = \left\langle \exp(i\mathbf{q}_1(\mathbf{r}_1-\mathbf{r}_0) + i\mathbf{q}_2(\mathbf{r}_3-\mathbf{r}_2))\right\rangle = \left\langle \exp(i\mathbf{q}_1\mathbf{R}_1 + i\mathbf{q}_2\mathbf{R}_2)\right\rangle, \quad (9)$$

where $\mathbf{R}_1 = \mathbf{r}_1 - \mathbf{r}_0$ is the spin displacement. Considering the cumulant expansion of the signal, in the absence of flow and keeping only terms up to the order $q^2$, the logarithm of the signal can be approximated by:

$$\log S(\mathbf{q}_1,\mathbf{q}_2) \approx -\frac{1}{2}\left\langle \phi^2 \right\rangle = -\frac{1}{2}(q_{1i}q_{1j} + q_{2i}q_{2j})\left\langle R_{1i}R_{1j}\right\rangle - q_{1i}q_{2j}\left\langle R_{1i}R_{2j}\right\rangle, \quad (10)$$

where we employed the Einstein summation convention, with the subscript letters labelling Cartesian components of vectors and tensors. The term $\left\langle R_{1i}R_{1j}\right\rangle$ represent the mean squared displacement and is equal to $2D_{ij}(\Delta)\Delta$ where $D_{ij}(\Delta)$ is the time dependent diffusion tensor which can be estimated based on SDE sequences. The second term in Eq. (10) $\left\langle R_{1i}R_{2j}\right\rangle$ is the displacement correlation tensor (Jespersen and Buhl 2011), which, as shown by Jespersen (Jespersen 2012), can be recovered by measuring the diffusion tensor at three different time points:

$$\left\langle R_{1i}R_{2j}\right\rangle = D_{ij}(2\Delta+\tau_m)\cdot(2\Delta+\tau_m) + D_{ij}(\tau_m)\cdot(\tau_m) - 2D_{ij}(\Delta+\tau_m)\cdot(\Delta+\tau_m) \quad (11)$$

Thus, the information measured by DDE sequences in the low q value regime can also be obtained from SDE measurements with different diffusion times.

From an experimental perspective, using the SDE and DDE phantom data from (Morozov, Bar et al. 2013), Jespersen showed that in the low q regime the estimated effective radii are indeed similar for SDE and DDE. Another recent study which compared the ability of SDE and DDE to estimate pore size distribution using a wide range of q-values (beyond the range where the cumulant expansion in Eq. (10) is sufficient), showed a good agreement between the two approaches in glass capillary phantoms and in ex-vivo spinal cord, although the authors reported a slightly better differentiation between white matter regions with DDE measurements.

When it comes to estimating axon diameters, the question of optimal gradient waveforms as well as the resolution limit, i.e. the smallest diameter which can be estimated given the hardware capabilities in terms of gradient strength and SNR, has been recently discussed in the literature (Drobnjak, Zhang et al. 2015, Ianus, Shemesh et al. 2017, Nilsson, Lasic et al. 2017, Kakkar, Bennett et al. 2018). Nevertheless, a systematic comparison of the SDE and DDE approaches for pore size estimation including practical aspects is still lacking.

## 3.2 Beyond $q^2$: estimation of pore size distribution

Besides estimation of an average pore size, DDE approaches can also be extended to estimate a distribution of sizes, which can more closely characterize the range of length scales in the tissue. Mapping the distribution can be done in either a parametric or non-parametric way. The advantage of assuming a certain parametric distribution (for instance log-normal or Gamma distributions are usually employed in the diffusion MRI literature (Assaf, Blumenfeld-Katzir et al. 2008, Sepehrband, Alexander et al. 2016)) is that it usually requires only one additional parameter to be fitted. Nevertheless, this might bias the estimates if the underlying size distribution is different. On the other hand, a non-parametric approach offers more flexibility. In this situation, the signal is usually written as a sum over signal contributions from compartments with different sizes ($S_n$) as well as additional terms if deemed necessary, e.g. for free diffusion ($S_{free}$):

$$S(\mathbf{q}_1,\mathbf{q}_2) = \sum_n f_n S_n(\mathbf{q}_1,\mathbf{q}_2,R_n) + S_{free}(\mathbf{q}_1,\mathbf{q}_2) \quad (12)$$

Then the signal fractions $f_n$ of the different compartments are fitted to minimize the difference between the measured and estimated signals. As discussed in Section 3.1.3, to capture information truly beyond SDE, modelling approaches also need to probe effects on the order of $q^4$ or higher. Estimating the signal fractions and the intra-compartmental diffusivity from Eq. 12 is a highly ill-posed problem, and different regularization approaches, for instance assuming a smooth distribution of sizes, have been proposed in the literature to stabilize the inversion (Benjamini and Nevo 2013, Benjamini, Komlosh et al. 2016, Methot, Ulloa et al. 2017).

### 3.2.1 Phantom validation

On preclinical scanners, several studies have shown the feasibility of employing DDE acquisitions to estimate a non-parametric distribution of pore sizes using micro-capillary phantoms (Benjamini, Katz et al. 2012, Benjamini and Nevo 2013, Benjamini, Komlosh et al. 2014, Morozov, Bar et al. 2015, Anaby, Morozov et al. 2019). Although the earlier studies (Benjamini, Komlosh et al. 2014) suggested experimental benefits of DDE over SDE for pore size distribution estimation, the more recent work by Anaby et al (Anaby, Morozov et al. 2019), which provides a fairer comparison between sequences, does not show an obvious advantage of DDE. As phantoms have proven a very useful tools for investigating dMRI contrasts and modelling techniques, an optimized design of physical phantoms tailored for pore size estimation is discussed in (Komlosh, Benjamini et al. 2017).

On a clinical scanner, a recent study by Duchene et al (Duchêne, Abarca-Quinones et al. 2020) has shown the feasibility of DDE sequences with short mixing time to estimate a distribution of pore sizes, which, in an asparagus phantom, closely corresponded to the values measured by light microscopy.

#### 3.2.2 Preclinical and clinical applications

In addition to estimating the mean pore size, several studies have focused on characterizing the distribution of pore sizes, based on DDE measurements. For instance, Benjamini et al (Benjamini, Komlosh et al. 2016) employed an optimized DDE protocol over a range of gradient strengths, diffusion times and relative angles to estimate the volume weighted and number weighted axon diameter distribution in the ferret spinal cord, using a similar approach to the one described in Eq. (12). The reconstructed distributions match well the histological values and an example is illustrated in Figure 3a. Comparing the performance of angular DDE and SDE protocols with multiple diffusion times for characterizing the distribution of axon diameters in ex-vivo porcine spinal cord, Anaby et al reported only subtle differences (Anaby, Morozov et al. 2019).

In a recent in-vivo study, Duchene et al employed a DDE protocol with parallel and anti-parallel gradients on a clinical scanner to analyse the pore size distribution and perfusion fraction of two tumour allografts in a rodent (Duchêne, Abarca-Quinones et al. 2020). The results showed significant differences between viable and necrotic areas, as well as a good correspondence with histological values.

### 3.3 High q-value regime: diffusion-diffraction patterns

The high q-value regime in diffusion NMR is also of great interest, and, in a seminal paper, Callaghan showed that the restricted diffusion signal measured with SDE exhibits diffraction patterns as q is increased (Callaghan, Coy et al. 1991). To illustrate this in a simple example, let's consider a SDE sequence ($\delta\to 0$, $\Delta\to\infty$) which is applied perpendicular to cylindrical fibres of radius R. In this case, Eq. (1) becomes

$$S = \left|\tilde{\rho}(\mathbf{q})\right|^2 \tag{13}$$

where $\tilde{\rho}(\mathbf{q}) = \int d\mathbf{r}\rho(\mathbf{r})\exp(-i\mathbf{qr})$ is the Fourier transform of the pore density function. For diffusion inside a homogeneous cylinder of radius R, the pore space function $\rho(\mathbf{r})$ is $(\pi R^2)^{-1}$ for r<R and 0 otherwise. Thus, calculating the reciprocal of the pore space function, we get:

$$\tilde{\rho}(\mathbf{q}) = \frac{2J_1(qR)}{qR}, \tag{14}$$

where $J_1$ is the first order Bessel function (Callaghan, Coy et al. 1991). As the signal vanishes when $J_1(qR) = 0$, the diffraction pattern can also inform on the restriction size.

The diffusion-diffraction patterns have also been investigated for DDE sequences (Ozarslan and Basser 2007, Shemesh, Ozarslan et al. 2010). To illustrate this in a simple example, let's consider a DDE sequence ($\tau_m$=0, $\Delta\to\infty$) with $\mathbf{q}_2 = -\mathbf{q}_1$ (i.e. the 2nd and 3rd pulses overlap and have the same orientation) which, as the SDE in the previous examples, is applied perpendicular to cylindrical fibres of radius R. In this case, Eq. (5) becomes

$$S = \tilde{\rho}(\mathbf{q})\tilde{\rho}(\mathbf{q})\tilde{\rho}(2\mathbf{q})^*. \tag{15}$$

Given the expression of $\tilde{\rho}(\mathbf{q})$ in Eq. (15), the signal vanishes when $J_1(qR) = 0$ or $J_1(2qR)=0$. Thus, the first zero crossing of the DDE signal happens when 2qR takes the value of the first $J_1$ zero, thus at half the q value, and implicitly half the gradient strength, compared to the first diffraction minimum for an SDE sequence.

Moreover, in case of a distribution of pore sizes, the zero crossings are still detectable for DDE sequences, which is not the case for SDE diffraction minima. This effect was shown both using simulations (Ozarslan and Basser 2007) and experimental data in a phantom with a distribution of capillary sizes (Shemesh, Ozarslan et al. 2010). This effect was explained by the crossing to the negative parts of the DDE signal, a feature which cannot occur in SDE, leading to a “smoothing” of the diffraction patterns in the presence of size distributions for the latter. Moreover, DDE measurements with vanishing mixing time which retain phase information (Eq. (7)), can be combined with SDE measurements to provide a direct measurement of pore shape, both for point-symmetric pores (Shemesh, F. et al. 2012, Kiselev and Novikov 2013) as well as for arbitrary pore shapes (Kuder and Laun 2012).

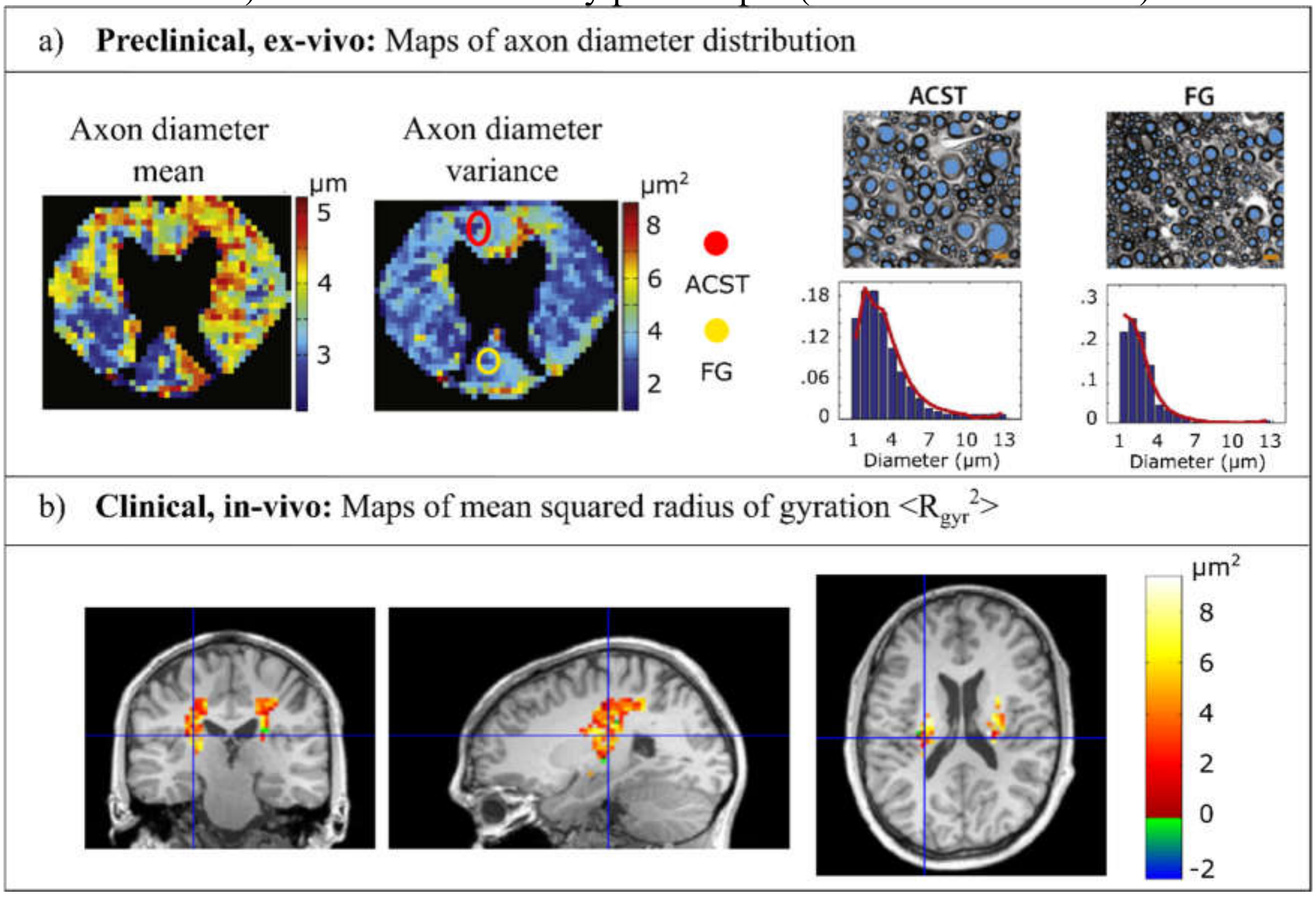

*Figure 3 a) Maps showing mean and variance of axon diameter distribution in ex-vivo ferret spinal cord and comparison between MRI and histology derived number-weighted axon diameter distributions in two ROIs (ACST – left anterior corticospinal tract; and FG - left gracile fasciculus). Figure adapted from Benjamini et al. (Benjamini, Komlosh et al. 2016) b) Maps showing the mean squared radius of gyration in the corticospinal tract, estimated from the angular dependence of DDE sequences with short mixing time. The negative values occur due to noise. Figure adapted from Koch and Finsterbusch (Koch and Finsterbusch 2011).*

# 4. Estimation of microscopic anisotropy

While pore size (mean and/or full distribution) estimation can be important for numerous applications, it is still a challenging task. Besides the usual confounding factors such as relaxation properties of the tissue, partial volume or noise, the estimated parameters are also influenced by the resolution limit due to hardware constraints and approximations of the biophysical models used to link the DDE signal modulation to the tissue geometry. On the other hand, microscopic anisotropy is a feature of heterogeneous systems that DDE can probe more directly, thus measuring it with less bias and assumptions.

Particularly, DDE with long mixing times have been extensively employed for quantifying microscopic diffusion anisotropy (μA) – a measurement of diffusion anisotropy independent of confounding orientation dispersion effects (Cory, Garroway et al. 1990, Mitra 1995, Cheng and Cory 1999, Callaghan and Komlosh 2002, Shemesh, Barazany et al. 2012). Moreover, DDE can be used to decouple different sources of non-Gaussian diffusion (Henriques et al., 2020). Importantly, situations in which significant orientation dispersion exists are prevalent in the CNS. In gray matter, neurites, dendrites, and astrocytic arms can be almost randomly oriented, while in white matter, the axons are typically dispersed along principal axes. In this section, we first provide the main theoretical considerations for estimating microscopic diffusion anisotropy from DDE measurements (section 4.1), then we discuss the theory and applications of rotationally invariant metrics of microscopic anisotropy (section 4.2), and, finally, we discuss how different non-Gaussian diffusion sources can be resolved using model-free DDE approaches (section 4.3).

## 4.1 Sensitivity to microscopic anisotropy and angular DDE

The sensitivity of DDE to microscopic diffusion anisotropy was conceptualized early on for randomly oriented ellipsoidal or spheroidal pores (Cory et al., 1990, Mitra 1995, Cheng and Cory 1999). In the long diffusion time $\Delta$ and long mixing time $\tau_m$ regimes, the DDE signal expression (Eq. (5) for an ensemble of pores (e.g., representing a complex microstructure within the voxel) can be written as:

$$S = \sum_n \left|\tilde{\rho}_n(\mathbf{q}_1)\right|^2 \left|\tilde{\rho}_n(\mathbf{q}_2)\right|^2 \tag{16}$$

Assuming perfectly randomly oriented anisotropic pores, the summation in Eq. (16) yields a signal which is invariant to the absolute gradient orientations $\hat{\boldsymbol{n}}_1$ and $\hat{\boldsymbol{n}}_2$, however, it depends on the relative angle $\varphi$ between the DDE gradients $\mathbf{q}_1$ and $\mathbf{q}_2$ (Mitra 1995). This angular dependence can be demonstrated by expanding Eq. (16) in a power series (Mitra 1995, Cheng and Cory 1999, Ozarslan 2009). Although early studies had expanded DDE signals using a Taylor series (Cheng and Cory 1999, Ozarslan 2009), more recent works employed a cumulant expansion instead, as it was shown to have a broader range of accuracy (Jespersen and Buhl 2011, Jespersen 2012). For DDE sequences with $|\mathbf{q}_1| = |\mathbf{q}_2| = q$, short gradient duration ($\delta \to 0$ ) and finite diffusion time $\Delta$, the cumulant expansion for randomly oriented pores up to the fourth order in $q$ is given by (Jespersen 2012):

$$\log S(\mathbf{q}_1, \mathbf{q}_2, \tau_m \to \infty) \approx -q^2 2\Delta D + \frac{1}{3} q^4 \Delta^2 D^2 W_{zzzz} + \frac{1}{4} q^4 \left[\cos^2 \varphi (Z_{zzzz} - Z_{zzxx}) + Z_{zzxx}\right], \tag{17}$$

where $\boldsymbol{W}$ is the kurtosis tensor (Jensen, Helpern et al. 2005) and $\boldsymbol{Z}$ is the fourth order correlation tensor, which in the long mixing time limit becomes the covariance tensor $\boldsymbol{C}$ (i.e. $\boldsymbol{Z} \to 4\ {}^{2}\boldsymbol{C}$). From Eq. (17), one can note that an angular dependence emerges at long mixing times is a 4th order effect, a property which also holds for the Taylor expansion (Cheng and Cory 1999, Lawrenz, Koch et al. 2009, Ozarslan 2009). The effect of different sequence and substrate parameters has been explored in previous studies using theoretical and numerical approaches, e.g. in (Ozarslan and Basser 2008, Koch and Finsterbusch 2009, Ozarslan 2009). Figure 2b shows an example of this angular dependence for DDE signals generated for three different substrates. One can note that this angular dependence exists only if pores are locally anisotropic in the plane defined by the two DDE wave vectors.

From the angular dependence of DDE signals at long mixing times, several descriptions for microscopic anisotropy have been proposed. Compartment shape anisotropy was defined in (Ozarslan 2009) and assumes known pore shape; compartment eccentricity was a more phenomenological description of the effect size of local anisotropy (Shemesh, Barazany et al. 2012); and the microscopic anisotropy metric in (Lawrenz, Koch et al. 2009) was measured via a tensor approach. The relationships between these metrics have been previously reviewed by (Jespersen, Lundell et al. 2013). Here, we follow the latest nomenclature consensus reported in (Shemesh, Jespersen et al. 2015), in which the term microscopic diffusion anisotropy is defined to be proportional to the variance of the single pore diffusion tensor eigenvalues (i.e., $\mu A^2 \equiv \frac{3}{5}\mathrm{var}(\lambda_i)$ (Ianus, Jespersen et al. 2018), where $\lambda_i$ are the eigenvalues of the diffusion tensor). From the tensor $\boldsymbol{Z}$ in Eq. (17), $\mu A^2$ can be computed as $(Z_{zzzz} - Z_{zzxx})/4\Delta^2$. For an ensemble of pores, microscopic diffusion anisotropy squared is an additive quantity, i.e. the net $\mu A^2$ for the ensemble is the sum of the microscopic diffusion anisotropies of its components.

#### 4.1.1 Phantom validation

Early NMR studies of DDE with long mixing time focused on signal differences between measurements with parallel and orthogonal gradients, and investigated samples such as randomly oriented elongated yeast cells (Cheng and Cory 1999), polydomain lyotropic liquid crystals (Callaghan and Komlosh 2002) and randomly oriented silica glass tubes (Komlosh, Horkay et al. 2007). In all of these, DDE's potential to resolve the microscopic anisotropy was clearly demonstrated; for example, Cheng and Cory differentiated between yeast cells that were irradiated – making them longer and more anisotropic on a local scale, despite being completely randomly oriented – from normal spherical yeast cells.

In later NMR experiments, the full $\cos^2\varphi$ angular dependence was explored in various samples. For instance, the signal measured in microcapillaries filled with water closely matches the theoretically curves predicted from diffusion restricted in cylindrical pores (Shemesh, Ozarslan et al. 2010). Subsequent studies performed angular DDE measurements to investigate toluene-in-water emulsion, quartz sand and spherical yeast cells (Shemesh, Adiri et al. 2011, Shemesh, Ozarslan et al. 2012). In short, these experiments successfully showed that the $\cos^2\varphi$ angular dependency is characteristic to specimens comprising of anisotropic microstructures.

#### 4.1.2 Preclinical and clinical applications of angular DDE

For biomedical imaging applications, angular DDE measurements have been first performed in NMR experiments to characterize microscopic anisotropy in fixed excised gray matter (Komlosh, Horkay et al. 2007, Shemesh and Cohen 2011) and white matter (Shemesh and Cohen 2011). For instance, Figure 4a shows the measured DDE signals as a function of different angles $\varphi$ for two biological specimens, where one can appreciate that the $\cos^2\varphi$ dependency is present for anisotropic gray matter and not for spherical yeast cells. Subsequent studies extended the angular DDE approach to imaging. On preclinical scanners, Komlosh et al (Komlosh, Lizak et al. 2008) investigated microscopic anisotropy in the fixed pig spinal cord while Shemesh and Cohen applied this approach to image the rat brain (Shemesh, Barazany et al. 2012). Lawrenz and Finsterbusch used a pig spinal cord to show that the DDE angular modulation can also be measured on a clinical scanner (Lawrenz and Finsterbusch 2011).

For in-vivo experiments, in a preclinical set-up, Shemesh and Cohen employed angular DDE experiments to study microscopic anisotropy in the rat brain (Shemesh, Barazany et al. 2012). In a clinical set-up, Lawrenz and Finsterbusch explored for the first time the DDE angular modulation in the human brain (Lawrenz and Finsterbusch 2013), and a later study by Avram et al extended the acquisition sequence to include four gradient pairs (Avram, Ozarslan et al. 2013) for a similar angular experiment.

These imaging studies(Lawrenz and Finsterbusch 2011, Shemesh, Barazany et al. 2012, Avram, Ozarslan et al. 2013, Lawrenz and Finsterbusch 2013), performed both on preclinical and clinical systems, showed regional variations in the DDE angular modulations that depend on the expected differences in tissue microstructure as well as on the plane spanned by the two gradient directions, an effect also illustrated in Figure 2c. These studies also provided some preliminary approaches to remove the effect of macroscopic anisotropy from the derived metrics and prompted the development of rotationally invariant acquisitions and data analysis strategies, which have been employed in more recent studies, as discussed in the next section.

## 4.2. Rotational invariant measures of microscopic anisotropy

The theoretical expressions presented in the previous section 4.1 assume that pores are randomly organized on the voxel space; however, tissue may include structures with different degrees of macroscopic anisotropy. For instance, in white matter, fibres usually have an orientation distribution pointing towards a primary direction. In this case, the $\cos^2\varphi$ angular modulation will also depend on the orientation of the plane spanned by the two gradient vectors (Ozarslan and Basser 2008, Lawrenz, Koch et al. 2009), as illustrated in Figure 2c for parallel cylinders. To decouple these conflating effects of macroscopic orientation coherence and microscopic tissue features, rotationally invariant $\mu A$ acquisition schemes – typically harnessing DDE signals averaged along different pairs of uniformly sampled directions – have been proposed (Jespersen, Lundell et al. 2013, Lawrenz and Finsterbusch 2013, Yang, Tian et al. 2018). For instance, first attempts to provide rotational invariant estimates of microscopic anisotropic were performed based on a tensor model (Lawrenz, Koch et al. 2009, Lawrenz and Finsterbusch 2013) or by incorporating information from the standard DTI model (Avram, Ozarslan et al. 2013). In a later study, Jespersen et al. (Jespersen, Lundell et al. 2013) proposed the use of direction pairs sampled based on the theory of numerical integration of polynomials which allows the exact calculation of averaged signals up to the 4th order in $q$. From this quadrature framework, rotationally invariant $\mu A$ estimates can be obtained from the directionally averaged signal of DDE measurements with parallel ($\bar{S}_{\parallel}$) and orthogonal ($\bar{S}_{\perp}$) gradient orientations (Jespersen et al., 2013):

$$\mu A^2 = \frac{\log(\bar{S}_{\parallel}) - \log(\bar{S}_{\perp})}{q^4\Delta^2}, \tag{18}$$

where, $\bar{S}_{\parallel} = \sum \frac{1}{N_{\parallel}} S_{\parallel}$ and $\bar{S}_{\perp} = \sum \frac{1}{N_{\perp}} S_{\perp}$ with $N_{\parallel}$ and $N_{\perp}$ the number of measurements with parallel and orthogonal gradients, respectively. Although the original direction scheme in (Jespersen, Lundell et al. 2013) consisted of 12 parallel and 60 orthogonal measurements, it is important to note that Eq. (18) is general to any quadrature framework that allows the estimation of averaged signals. For example, assuming Gaussian diffusion in each anisotropic pore, sampling schemes with smaller number of measurements have been proposed to promote clinical translation (Yang, Tian et al. 2018) (Lundell, Dyrby et al. 2015, Kerkelä, Henriques et al. 2020). On the other hand, other sampling schemes have been proposed to incorporate a higher number of parallel and orthogonal DDE measurements to improve the robustness of the estimates (Coelho, Pozo et al. 2019, Henriques, Jespersen et al. 2020 ). To account for finite gradient duration, Eq. (18) can be further written in terms of the b value of each gradient pair (Ianus, Jespersen et al. 2018).

From $\mu A$ estimates, it may be useful to define a normalized metric of microscopic diffusion anisotropy which is independent of the system's mean diffusivity D (Jespersen, Lundell et al. 2013, Jespersen, Lundell et al. 2014):

$$\mu FA = \sqrt{\frac{3}{2}\frac{\mu A^2}{\mu A^2 + \frac{3}{5}D^2}} \tag{19}$$

This normalized metric (μFA) has been termed microscopic fractional anisotropy and ranges from 0 in isotropic pores to 1 in highly anisotropic pores, similarly to the fractional anisotropy (FA) obtained from DTI. However, as opposed to FA, μFA is not conflated by fibre orientation effects, and represents the microscopic anisotropy in the system.

It is important to note that Eq. (17) and (18) only consider effects up to the 4th order of $q$ and the derived $\mu A$ estimates can be biased for larger $q$ values (Ianus, Jespersen et al. 2018). To minimize the bias due to higher order terms, Ianus et al. (2018) proposed the estimation of $\mu A$ from averaged DDE parallel and orthogonal signals acquired at multiple b-values using the following polynomial fit:

$$\log(S_{\|}{}^{a}(q)) - \log(S_{\perp}{}^{a}(q)) = q^4\Delta^2\mu A^2 + P_3 q^6 \Delta^3 \tag{20}$$

where $P_3$ is a term used to capture higher order contributions up to the 6th order in $q$. Although this strategy may improve the accuracy of $\mu A$ estimates, incorporating higher order terms in polynomial fits is well known to impact the precision of parameter estimates (Chuhutin, Hansen et al. 2017, Kiselev 2017). As an attempt to improve precision and accuracy, recent studies have estimated $\mu A$ based on the analytical expression of higher order coefficients assuming a substrate of randomly oriented identical pores (Shemesh, Rosenberg et al. 2017, Ianus, Jespersen et al. 2018, Najac, Lundell et al. 2019). Nevertheless, this approach might also introduce a bias if the model assumptions do not hold (Ianus, Drobnjak et al. 2016). More implications of DDE measurements for microstructure modelling will be discussed in Section 5.

#### 4.2.1 Preclinical and clinical applications of rotationally invariant $\mu A$ metrics

Building on earlier angular DDE studies (Section 4.1.2), the more recent applications of DDE for mapping microscopic diffusion anisotropy have been focused on methods which estimate rotationally invariant metrics of $\mu A$.

In a preclinical set-up, Jespersen at al (Jespersen, Lundell et al. 2013) tested and applied $\mu A$ metrics, that were derived from the cumulant expansion of the signal measured following the 5-design gradient scheme (Eq. (18) - (19)), to study microscopic anisotropy in a normal Vervet monkey brain, ex-vivo. The results showed that indeed the derived $\mu A$ metrics are robust to the rotation of the gradient directions and are not conflated by the macroscopic orientation distribution of the fibres. This effect is illustrated in Figure 4a, which shows the comparison between standard DTI fractional anisotropy maps and microscopic anisotropy maps from (Jespersen et al., 2013) – while FA maps show hypo-intensities in WM and GM regions (red and blue arrows, respectively) known for dispersed fibres, the more homogenous contrast in $\mu FA^2$ maps indicates lower sensitivity to confounding orientation effects. In a more recent ex-vivo experiment imaging the mouse brain, Ianus et al (Ianus, Jespersen et al. 2018) employed simulations and DDE measurements following the 5-design to study the b-value and diffusion time dependence of $\mu A$ metrics. The results suggested that simple tissue models such as infinite cylinders are not enough to explain the measured trends, and more complex models, for instance, incorporating structure along the fibre direction are required. A similar acquisition protocol has also been used for imaging the rat spinal cord, in-vivo, to characterize the tissue microstructure changes after spinal cord injury (Budde, Skinner et al. 2017).

In a clinical set-up, Lawrenz and Finsterbusch (Lawrenz and Finsterbusch 2015) have employed DDE with parallel and orthogonal gradients to map the microscopic anisotropy in the human white matter based on rotationally invariant metrics derived from the Taylor expansion of the signal. Moreover, they have investigated different sequence preparations and gradient direction schemes, showing, both in a water phantom and in-vivo, that including reversed-order and antiparallel direction combinations can help to reduce the unwanted influence of experimental imperfections, imaging artifacts and/or residual restriction effects at short mixing time (Lawrenz and Finsterbusch 2015). Specific aspects related to the sequence implementation will be discussed in detail in Section 8. In a more recent application to study age-related changes in the human brain, Lawrenz et al (Lawrenz, Brassen et al. 2016) showed that microscopic anisotropy measures are more sensitive to degenerative processes than DTI's FA measures, especially in white matter regions that contain crossing bundles. In addition to characterizing WM, Lawrenz and Finsterbusch have also shown the feasibility of mapping microscopic anisotropy in the human cortex, in-vivo (Lawrenz and Finsterbusch 2019).

Rotationally invariant microscopic anisotropy estimates have also been applied to characterize abnormal alteration in human brain, such as WM multiple sclerosis lesions (Yang, Tian et al. 2018, Andersen, Lasič et al. 2020). Yang et al showed that μFA hypo-intensities could better depict WM lesions compared to standard FA, that also had low values in WM regions featuring fibre dispersion and crossing. The results for a representative patient from this study are illustrated in Figure 4b.

In the past, several techniques have been also proposed to estimate microscopic diffusion anisotropy based on SDE measurements (Callaghan, Jolley et al. 1979, Kaden, Kelm et al. 2016, Kaden, Kruggel et al. 2016). Nevertheless, such approaches make strong assumptions about the underlying tissue microstructure, for instance that it comprises of identical pores, which can result in biased $\mu A^2$ values (Lampinen, Szczepankiewicz et al. 2017, Henriques, Jespersen et al. 2019). On the other hand, methods that employ advanced sequence that vary the gradient orientation within one measurement, such as the DDE approaches discussed in this section or techniques which employ different b-tensor shapes, can provide estimates of microscopic diffusion anisotropy without confounding effects from a distribution of pore sizes, which is the case in biological tissue. B-tensor methods have also seen an increasing interest in recent years, for instance, to characterize the heterogeneity of brain microstructure (Westin, Knutsson et al. 2016), kidney tissue (Nery, Szczepankiewicz et al. 2019), discriminate between tumour types and grades (Szczepankiewicz, van Westen et al. 2016, Nilsson, Szczepankiewicz et al. 2020), etc.

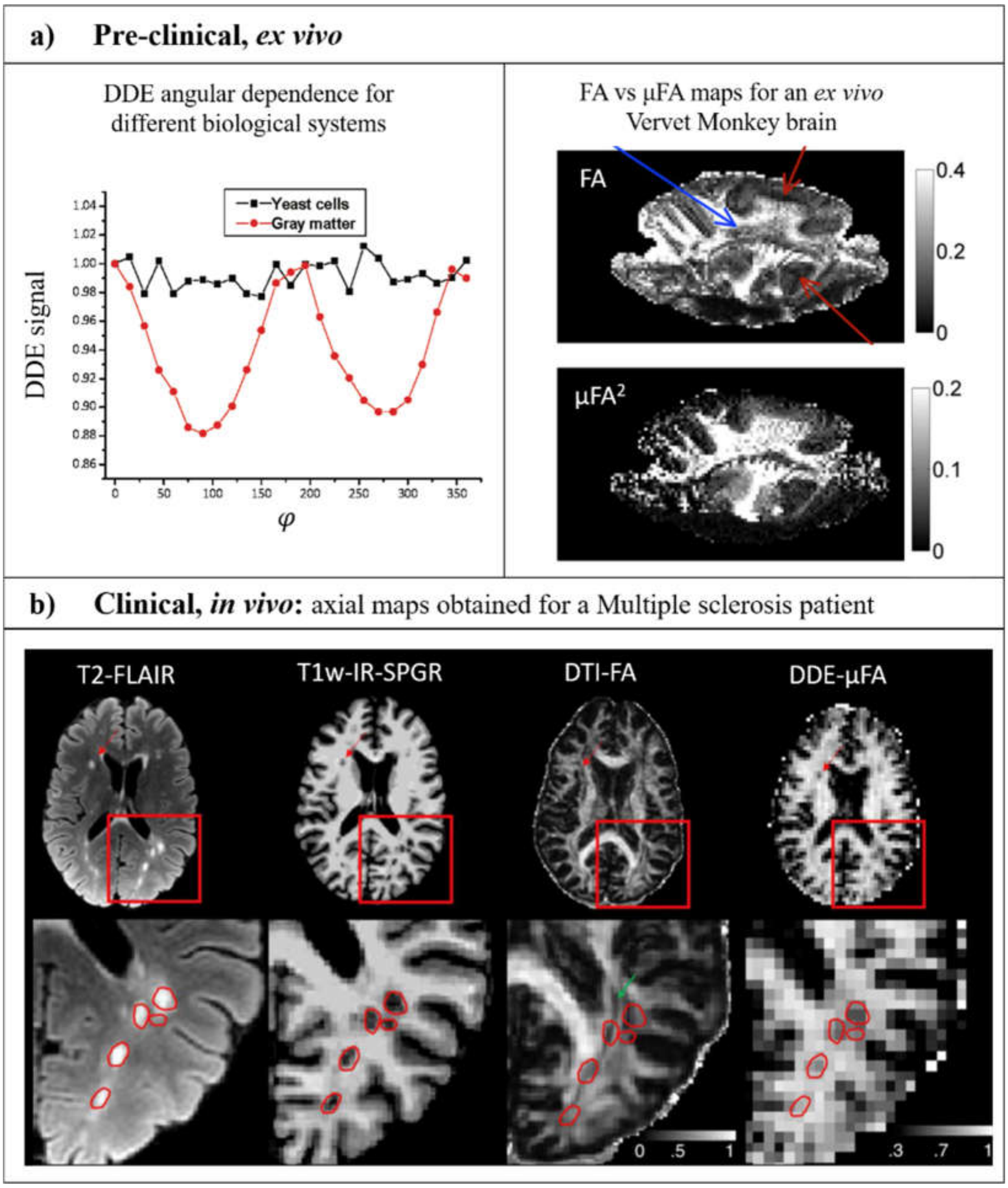

*Figure 4* ***Characterization of microscopic anisotropy using DDE experiments in a) preclinical and b) clinical settings. a), left panel:*** *DDE angular dependency for isotropic yeast cells and anisotropic grey matter cells measured on a 8.4T Bruker NMR spectroscopic system (panel adapted from (Shemesh, Barazany et al. 2012));* ***a), right panel:*** *FA and μFA² rotational invariants maps for a sagittal slice of an ex vivo Vervet monkey brain obtained on Varian 4.7 T animal scanner (panel adapted from (Jespersen, Lundell et al. 2013)) – red and blue arrows indicate FA hypo-intensities in gray and white matter regions respectively;* ***c)*** *Representative brain axial slice images for T2-FLAIR, T1-weighted, FA and μFA of an multiple-sclerosis patient obtained on 3T clinical scanner (panel adapted from (Yang, Tian et al. 2018)).*

## 4.3 Mapping the Correlation Tensor using DDE MRI

Further to mapping microscopic anisotropy, it has been recently shown that DDE acquisitions with long mixing times can disentangle the main sources influencing non-mono-exponential diffusion-driven signal decays. It is already well known from SDE literature that quantifying the total degree of non-Gaussian diffusion can be highly useful, e.g., for increasing the sensitivity of the experiment towards disease (Falangola, Jensen et al. 2008, Cheung, Wang et al. 2012 , Rudrapatna, Wieloch et al. 2014, Sun, Chen et al. 2015). Indeed, diffusional kurtosis tensor, which can be extracted from diffusion-weighted signals measured at different b-values (Jensen, Helpern et al. 2005), has played an important role in biomedical imaging. Nevertheless, in heterogeneous samples, the rotational invariants of the kurtosis tensor conflate multiple effects (Paulsen, Ozarslan et al. 2015). For instance, considering SDE signals averaged across uniformly sampled directions, the total measured kurtosis $K_T$ is the sum of three different sources (Henriques, Jespersen et al. 2020), as illustrated in Figure 5a:

$$K_T = K_{aniso} + K_{iso} + K_{intra} \tag{21}$$

where $K_{aniso}$ rises due to microscopic anisotropy ($K_{aniso} = 2\mu A^2/D^2$), $K_{iso}$ is associated with the variances of mean diffusivities across different compartments ($K_{iso} = 3var(D^c)/D^2$, with $D^c$ the mean diffusivity of each individual compartment $c$), and $K_{intra}$ is a weighted sum of the non-Gaussian effects of individual restricted compartments ($K_{intra} = (D^c)^2 K^c / D^2$, with $K^c$ the mean kurtosis of each individual compartment $c$). This latter term is referred to as intra-compartmental kurtosis (Henriques, Jespersen et al. 2020). Note that these three kurtosis sources cannot be decoupled with SDE experiments.

Several studies have employed multidimensional encoding using continuous gradient waveforms to resolve $K_{aniso}$ and $K_{iso}$ by subtracting the information provided by microscopic anisotropy from the total measured kurtosis (e.g. (Lasic, Szczepankiewicz et al. 2014, Szczepankiewicz, Lasic et al. 2014, Westin, Knutsson et al. 2016, Topgaard 2017)). However, some recent studies have pointed out that such approaches may be biased by effects of diffusion time dependence and neglected intra-compartmental kurtosis sources (de Swiet and Mitra 1996, Jespersen, Olesen et al. 2019). First insights that $K_{intra}$ could be measured from DDE experiments were provided by Paulsen and colleagues (Paulsen, Ozarslan et al. 2015) This study showed that for isotropic pores or well-aligned anisotropic pores, $K_{intra}$ could be measured from the frequency dependence of a symmetrized DDE experiments (performed with varying $\mathbf{q}_1$ and $\mathbf{q}_2$ magnitudes while maintaining $\mathbf{q}_1$ and $\mathbf{q}_2$ directions fixed and $|\mathbf{q}_1+\mathbf{q}_2|$ constant). This approach was first applied on a glass capillary phantom and an asparagus sample in NMR experiments (Paulsen, Ozarslan et al. 2015), and later on a three compartmental phantom and in-vivo rat brain in a preclinical MRI system (Ji, Paulsen et al. 2019). The method used in these studies, however, conflates $K_{intra}$ with mesoscopic dispersion effects (Paulsen, Ozarslan et al. 2015).

More recently, Henriques et al. developed a general and accurate scheme for resolving the kurtosis sources – an approach referred to as the correlation tensor imaging (CTI) (Henriques, Jespersen et al. 2020). CTI is based on the correlation tensors measured from the cumulant expansion of DDE signals in the long mixing time regime. In particular, for any given directions and magnitudes of $\mathbf{q}_1$ and $\mathbf{q}_2$, the DDE signals up to the fourth order in $q$, can be expressed as (Jespersen 2012):

$$\begin{aligned} \log S(\mathbf{q}_1, \mathbf{q}, \ \tau_m \to \infty) \approx & -\left(q_{1i}q_{1j} + q_{2i}q_{2j}\right)\Delta D_{ij} \\ & + \frac{1}{6}\left(q_{1i}q_{1j}q_{1k}q_{1l} + q_{2i}q_{2j}q_{2k}q_{2l}\right)\Delta^2 D^2 W_{ijkl} \\ & + \frac{1}{4} q_{1i}q_{1j}q_{2k}q_{2l} Z_{ijkl} \end{aligned} \tag{22}$$

where $\boldsymbol{W}$ is the kurtosis tensor (Jensen, Helpern et al. 2005) and $\boldsymbol{Z}$ is the fourth order correlation tensor, which in the long mixing time limit becomes the covariance tensor $\boldsymbol{C}$ (i.e. $\boldsymbol{Z} \to 4\ {}^2\boldsymbol{C}$). The tensors $\boldsymbol{D}$, $\boldsymbol{W}$ and $\boldsymbol{Z}$ provide sufficient information to disentangle all three kurtosis sources (in the absence of flow and exchange): 1) $K_{aniso}$ and $K_{iso}$ can generally be estimated from $\boldsymbol{D}$ and $\boldsymbol{Z}$ since they do not depend on diffusion time and wall-reflection effects; then 2) $K_{intra} =$ $K_T$   $K_{aniso}$   $K_{iso}$ can be estimated by subtracting $K_{aniso}$ and $K_{iso}$ from the total kurtosis extracted from $\boldsymbol{W}$. Note that to decouple $\boldsymbol{Z}$ from $\boldsymbol{W}$, Eq. (22) has to be fitted to DDE data acquired with different direction and magnitude combinations of $\mathbf{q}_1$ and $\mathbf{q}_2$ (Henriques, Jespersen et al. 2020).

So far, CTI has been applied to in vivo and ex vivo rodent brains, as illustrated in Figure 5b. As expected, CTI's $K_{aniso}$ had larger values in areas with high microscopic anisotropy variances, such as WM (white arrows in Figure 5b), while $K_{iso}$ had larger values in regions with increased free water partial volume effect (red arrows in Figure 5b). In contrast to the estimates from (Ji, Paulsen et al. 2019) which were conflated by the orientation distribution, the $K_{intra}$ estimates provided by CTI revealed the absence of high intensities of intra-compartmental kurtosis in white matter and non-zero positive values in gray matter.

Although the separation of kurtosis sources is promising for the characterization of microstructural properties in a model-free manner (which bodes well for many future applications), it is important to note that further studies are required to relate the sensitivity and specificity of $K_{intra}$ estimates to microstructural properties, and to mitigate possible confounding factors of CTI's framework, such as biases from higher-order-terms not considered in the cumulant expansion truncation.

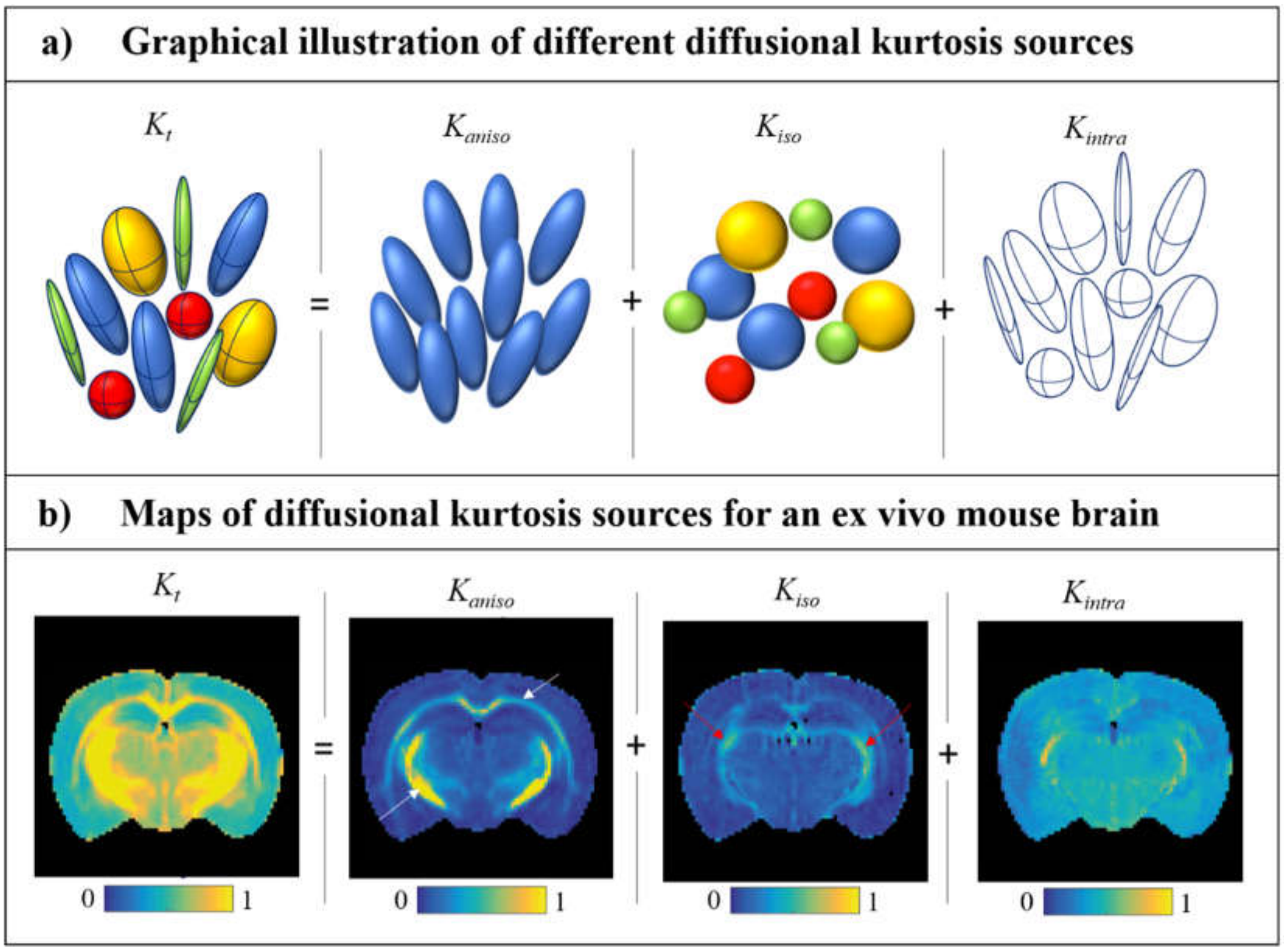

*Figure 5 Different diffusional kurtosis sources. a) Graphical representation of different diffusional kurtosis sources: the total kurtosis ($K_t$) may be decomposed into three kurtosis sources using DDE based techniques: anisotropic kurtosis ($K_{aniso}$) which arises due to ensemble microscopic anisotropy (diffusion variance over different directions); isotropic kurtosis ($K_{iso}$) which arises due to the variance of compartments' mean diffusivities and intra-compartmental kurtosis ($K_{intra}$) which depends on restricted diffusion effects (represented in the figure by the walls of each compartment). b) Maps of different diffusional kurtosis sources obtained using CTI (panel adapted from Henriques et al., 2020): total kurtosis, anisotropic kurtosis (white matter regions with high microscopic anisotropy are indicated by the white arrows), isotropic kurtosis (regions containing high partial volume effect with ventricles are pointed out by the red arrows), and intra-compartmental kurtosis.*

# 5. Implications for microstructural modelling

In this section, we will review how the unique information available in DDE can be useful for both microstructural model fitting and validation. As an example, we will use the widely used family of models referred to as the "standard model" in diffusion microstructure imaging (Novikov, Kiselev et al. 2018, Novikov, Fieremans et al. 2019).

## 5.1 The standard model for diffusion in white matter

Briefly, the standard model (SM) represents diffusion in white matter in terms of a sum of non-exchanging Gaussian diffusion populations, representing different microscopic compartments (Novikov, Kiselev et al. 2018, Novikov, Fieremans et al. 2019). A characteristic feature is the approximation of axons as zero radius cylinders, or sticks, due to their small diameter of typically ~0.2-5 μm. Thus, for a fibre segment with parallel axons oriented along the direction $\hat{u}$, the normalized diffusion weighted signal is given by (Jespersen, Kroenke et al. 2007, Jespersen, Bjarkam et al. 2010):

$$K(b,\hat{\boldsymbol{u}}) = f\exp\left[-D_a b_{ij}u_i u_j\right] + (1-f)\exp\left[-bD_e^{\perp} - (D_e^{\parallel} - D_e^{\perp})b_{ij}u_i u_j\right] \qquad (23)$$

where $f$ is the $T_2$-weighted (T1-, proton-, etc, weighted) stick volume fraction, $D_a$ is the intra-neurite diffusivity, $D_e^{\parallel}$ is the extra-neurite axial diffusivity, and $D_e^{\perp}$ is the extra-neurite radial diffusivity. Note that some studies incorporate a component with an isotropic constant diffusivity of 3 $\mu m^2/ms$ to model free water contributions (Zhang, Schneider et al. 2012, Reisert, Kellner et al. 2017, Reisert, Kiselev et al. 2019), whereas others add a so-called dot compartment (Stanisz, Szafer et al. 1997, Kroenke, Bretthorst et al. 2006, Panagiotaki, Schneider et al. 2012), representing water which is trapped in small spaces and therefore has approximately $D \approx 0$. In general, one may need to add additional compartments to Eq. (23), but here we keep only the two most common compartments, intra-and extra-axonal spaces, for the sake of simplicity. To capture the signal from all fibre segments in the voxel, we introduce the fibre orientation distribution function (fODF) $\mathcal{P}(\hat{u})$ specifying the fraction of fibres along a particular direction. The fODF can be modelled using specific functions such as the Watson distribution (Jelescu, Veraart et al. 2016, Coelho, Pozo et al. 2019) or represented quite generally by its spherical harmonics decomposition (Tournier, Calamante et al. 2004, Anderson 2005). The advantage of the former is simplicity and analytical tractability, whereas the latter is more general and factorizes the signal after Laplace transformation. The total signal thus becomes

$$S(\mathrm{b}) = S_0 \int_{\mathbb{S}^2} \mathcal{P}(\hat{u})\mathrm{K}(\mathrm{b},\hat{u})d\hat{u} \qquad (24)$$

Note that, although here we focus on SDE and DDE, where $\mathrm{b} = b\hat{\boldsymbol{n}} \otimes \hat{\boldsymbol{n}}$ and $\mathrm{b} = b_1\hat{\boldsymbol{n}}_1 \otimes \hat{\boldsymbol{n}}_1 + b_2\hat{\boldsymbol{n}}_2 \otimes \hat{\boldsymbol{n}}_2$, respectively, Eq. (23) and (24) apply generally for any b-tensor (Topgaard 2017). Estimating the model parameters of the standard model by fitting to experimentally acquired data is therefore a promising avenue for estimating properties of living tissue on the cellular scale. However, before becoming adopted as the standard clinical tool, the model must be validated, and robust strategies for parameter estimation must be defined. In the following sections, we review these two aspects in the context of DDE.

## 5.2 Model parameter estimation

Several recent studies show that even when tissue models with a small number of compartments are considered, fitting SDE signals is ill-conditioned and its parameter estimates suffer from low accuracy and precision (Jelescu, Veraart et al. 2016, Novikov, Veraart et al. 2018) unless very high diffusion weighting is applied (Jespersen, Kroenke et al. 2006, Jespersen, Bjarkam et al. 2010). In the more typical regime of diffusion weighting achievable on clinical scanners however, most diffusion signals are well described by the cumulant expansion to order $b^2$ (DKI). At this level, the general two compartment standard model requires at least 18 parameters (4 kernel parameters and 14 fODF parameters). Although the diffusion and kurtosis tensors combined have 22 degrees of freedom, it was shown by Novikov et al. (Novikov, Veraart et al. 2018) that they do not uniquely determine the 18 SM parameters: the DKI parameters contain interdependent combinations of the SM parameters such that the fODF is overdetermined and the kernel underdetermined. In particular, two one-dimensional manifolds (branches) in the 18 dimensional SM parameter space give rise to exactly the same diffusion and kurtosis tensors, and thus practically indistinguishable signals at low to moderate diffusion weighting. Enforcing e.g. a Watson distribution for the fODF reduces each of these two branches to two separate point solutions (Jespersen, Olesen et al. 2018), corresponding roughly to one scenario in which the intra-axonal diffusivity is larger than the extra axonal axial diffusivity, and the opposite scenario with intra-axonal diffusivity being the smaller of the two. The choice between the two still requires independent information. However, as proved recently by Coelho et al (Coelho, Pozo et al. 2019) and by Reisert et al (Reisert, Kiselev et al. 2019), DDE in principle supplies enough information to resolve the degeneracy without the need for further assumptions, such as the Watson distribution. This can be appreciated by the fact that SDE, having $b_{ij} \propto n_i n_j$ only probes the totally symmetric part of the Z tensor in Eq. (17), while DDE measurements couple to a larger part of Z: for example, the contributions to the DDE signal of $Z_{xzxz}$ and $Z_{xxzz}$ can be independently varied, as also necessary for the determination of microscopic anisotropy. While all elements of Z tensor can thus be probed by the combination of SDE and DDE, triple diffusion encoding (TDE) (or spherical diffusion encoding (STE)) with $b_{ij} \propto \delta_{ij}$ (coupling to $Z_{ijij}$) and SDE (coupling to totally symmetric part of Z) alone or combined do not allow terms such as $Z_{xzxz}$ to be determined (Coelho, Pozo et al. 2019, Reisert, Kiselev et al. 2019), indicating that more information is captured with planar over spherical encoding. In agreement with this, optimizing the diffusion weighting waveform for SM parameter determination, Coelho et al identified a combination of linear and planar diffusion encoding on two shells (Coelho, Pozo et al. 2019). On the other hand, Lampinen et al found similar performance for protocols combining linear and spherical diffusion encoding with variable echo time (Lampinen, Szczepankiewicz et al. 2020), similarly to Afzali et al using simulations (Afzali, Tax et al. 2019). In all cases considered, protocols combining linear encoding such as SDE with planar encoding such as DDE performed better in terms of precision of model parameter estimates than SDE alone (Afzali, Tax et al. 2019). As an example, Dhital et al. used planar diffusion encoding to estimate intra-axonal diffusivity of SM (Dhital, Reisert et al. 2019), by using the planar diffusion weighting to suppress signal from compartments with spins that are mobile in all directions, such as extra axonal spins. Jensen and Helpern similarly used high diffusion weighting and TDE to determine both intra-axonal diffusivity and axonal water signal fraction (Jensen and Helpern 2018).

Therefore, in addition to the model free estimates of pore size and microscopic anisotropy (vide supra), the extra information captured by DDE can be used to resolve the degeneracies of the two-compartmental model or provide means to validate previous model constraints.

## 5.3 Model Validation

Despite its crucial role, model validation of diffusion MRI is inherently difficult. Other modalities, such as optical microscopy, used for comparison typically are highly invasive, have very small field of view, or are sensitive to other aspects of tissue microstructure. Under certain circumstances, diffusion weighting can provide some validation by itself. For example, the SM predicts that at high b-values, the powder averaged SDE signal should decay as a power law $b^{-1/2}$, which was recently verified (Veraart, Fieremans et al. 2019). The DDE has similarly been predicted to scale as $b^{-1}$ for SM (Herberthson, Yolcu et al. 2019), consistent with subsequent measurements (Afzali, Aja-Fernández et al. 2020). If sufficiently strong gradients are available, such that SDE can provide a reasonable estimation of SM parameters, DDE can be used as an independent validation, as was done in (Henriques, Jespersen et al. 2019) in fixed mouse brain. The authors found that the two compartment SM was not fully compatible with both SDE and DDE data; however, only powder averaged data were used to estimate model parameters, which may have affected the precision. Another possibility is the existence of additional compartments, such as the previously mentioned dot compartment or restricted diffusion inside cell bodies, which are present in small, but not negligible volume fractions in WM (typically 5-15%) (Stanisz, Szafer et al. 1997, Kroenke, Bretthorst et al. 2006, Panagiotaki, Schneider et al. 2012, Palombo, Ianus et al. 2020). The existence of the dot compartment in living human brain was also examined using spherical diffusion encoding by Dhital et al in (Dhital, Kellner et al. 2018), who showed that it accounted for less than 2% of the total signal. This was consistent with findings by Tax et al. (Tax, Szczepankiewicz et al. 2020), except in the cerebellum, where a slow diffusing signal fraction of ~10% was found. Although it has not been investigated yet how much cell body contribution can bias the SM estimates in WM, a recent work has shown that cell bodies have indeed a non-negligible contribution to the overall signal at high b values, in both WM and GM (Palombo, Ianus et al. 2020).

Others have attempted to identify possible constraints on the SM model parameters that might allow SDE data to unambiguously determine all independent parameters. Models of the neurite orientation dispersion and density imaging (NODDI) and spherical mean technique (SMT) are notable examples, where e.g. relations among intra-and extra-axonal diffusivities are imposed (Zhang, Schneider et al. 2012, Kaden, Kelm et al. 2016, Kaden, Kruggel et al. 2016, Tariq, Schneider et al. 2016). Such assumptions can also be tested with multidimensional encoding, for instance Lampinen et al (Lampinen, Szczepankiewicz et al. 2017) used spherical tensor encoding and echo time variation, and Henriques et al (Henriques, Jespersen et al. 2019) used DDE to compare the resulting $\mu A$ metrics. The additional DDE / STE data was not fully compatible with the original constraints, nevertheless, it can be used to improve the model assumptions (Guerreri, Szczepankiewicz et al. 2018).

# 6. DDE for correlation, exchange and filter studies

The previously described DDE techniques have mainly focused on experiments where the first and second gradient amplitudes are typically equal. However, the DDE parameter space is quite large, and spanning different gradient wavevectors, diffusion times, gradient durations and/or mixing times, opens up the possibility of multidimensional experiments, which can reveal new features of the tissue microstructure, composition and dynamics. (Callaghan and Xia 1991, Callaghan and Manz 1994, Callaghan and Komlosh 2002, Callaghan and Furo 2004).

## 6.1 Diffusion-Diffusion Correlation Spectroscopy (DDCOSY)

The DDCOSY approach (Godefroy and Callaghan 2003, Qiao, Galvosas et al. 2005), which aims to probe the displacement correlation in different directions, employs DDE acquisitions with short mixing time in which the amplitude of each gradient pair is stepped independently to obtain a two-dimensional measurement space, although in anisotropic samples the measurement space can be extended to include different gradient directions as well (Zong, Ancelet et al. 2017). The DDE data is analysed using a 2D Inverse Laplace Transform (ILT) to obtain diffusion-diffusion correlation maps. Then, comparing the maps obtained from DDE sequences with parallel and orthogonal gradient directions can inform on microscopic diffusion anisotropy. To illustrate this effect, we consider an anisotropic, axially symmetric microdomain with parallel and perpendicular diffusivities $D_{\|}$ and $D_{\perp}$. Then, the apparent diffusion coefficient along the direction of the diffusion gradient **G** depends on the angle θ between this direction and the main axis of the microdomain (Callaghan, Jolley et al. 1979):

$$D_{\theta} = D_{\|} \cos(\theta)^2 + D_{\perp} \sin(\theta)^2 \tag{25}$$

For a substrate consisting of randomly oriented anisotropic microdomains, the signal for a DDE sequence with parallel gradients along the (arbitrarily chosen) z direction is:

$$S_{zz}(b_1, b_2) = \frac{1}{2}\int_0^{\pi} \exp\left(-(b_1 + b_2)(D_{\|}\cos^2\theta + D_{\perp}\sin^2\theta)\right)\sin\theta d\theta, \tag{26}$$

and for a DDE sequence with orthogonal gradients (along z and x axis), the signal is:

$$\begin{aligned} S_{zx}(b_1, b_2) = \frac{1}{4\pi}\int_0^{2\pi}\int_0^{\pi} \exp[-b_1(D_{\|}\cos^2\theta + D_{\perp}\sin^2\theta) - \\ -b_2(D_{\|}\sin^2\theta\cos^2\phi + D_{\perp}\sin^2\phi + D_{\perp}\cos^2\theta\cos^2\phi)]\sin\theta d\theta d\phi. \end{aligned} \tag{27}$$

Thus, the 2D diffusion spectra obtained by ILT from data acquired with parallel and orthogonal gradients will have different appearances in the presence of microscopic anisotropy, as illustrated in Figure 6b and Figure 6c which simulate a DDCOSY experiment for randomly anisotropic Gaussian domains with $D_{\|} = 1$ μm²/ms and $D_{\perp} = 0.1$ μm²/ms. In the case of parallel gradients, the spectrum exhibits only diagonal peaks, while in the case of orthogonal gradients, the spectrum has a wide range of off-diagonal peaks, which reflect the 2D diffusion-diffusion probability distribution.

In macroscopically anisotropic substrates, the DDCOSY spectra will also depend on the choice of gradient directions. An in-depth analysis of various substrate configurations is presented in a recent study by Zong et al (Zong, Ancelet et al. 2017) which aims to extract an overall measure of microscopic anisotropy from the 2D spectra.

### 6.1.2 Preclinical applications

DDCOSY techniques have been first applied to characterize the microscopic structure of materials such as a polydomain lyotropic liquid crystal system, Aerosol OT–water (Callaghan and Komlosh 2002, Callaghan and Furo 2004), water and oil dynamics in food and micro-emulsion systems (Godefroy and Callaghan 2003), as well as in plant tissue (Qiao, Galvosas et al. 2005, Zong, Ancelet et al. 2017).

For biomedical applications, Zong et al employed 2D diffusion-diffusion correlation maps as well as a measure of fractional anisotropy derived from DDCOSY acquisitions with several combinations of gradient orientations to study diffusion properties of both healthy and tumour-bearing mouse brains, ex-vivo (Zong, Ancelet et al. 2017). To significantly shorten the acquisition time of 2D methods, Benjamini and Baser proposed a marginal distribution constrained optimization (MADCO) approach to obtain the 2D correlation function from a 1 D sampling and only several points in the 2D space (Benjamini and Basser 2016). This approach has been employed in a recent study to characterize the joint water mobility distributions perpendicular and parallel to the spinal cord axis in ex-vivo tissue (Benjamini, Hutchinson et al. 2020). As illustrated in Figure 7a, the study showed six diffusion spectral components with different microscopic anisotropy and water mobility. Moreover, one of the microenvironments was associated with injury-induced axonal degeneration, with reduced parallel diffusivity and increased perpendicular diffusivity, showing the potential of 2D diffusion correlation measurements to improve the specificity towards tissue microstructure compared to 1D techniques.

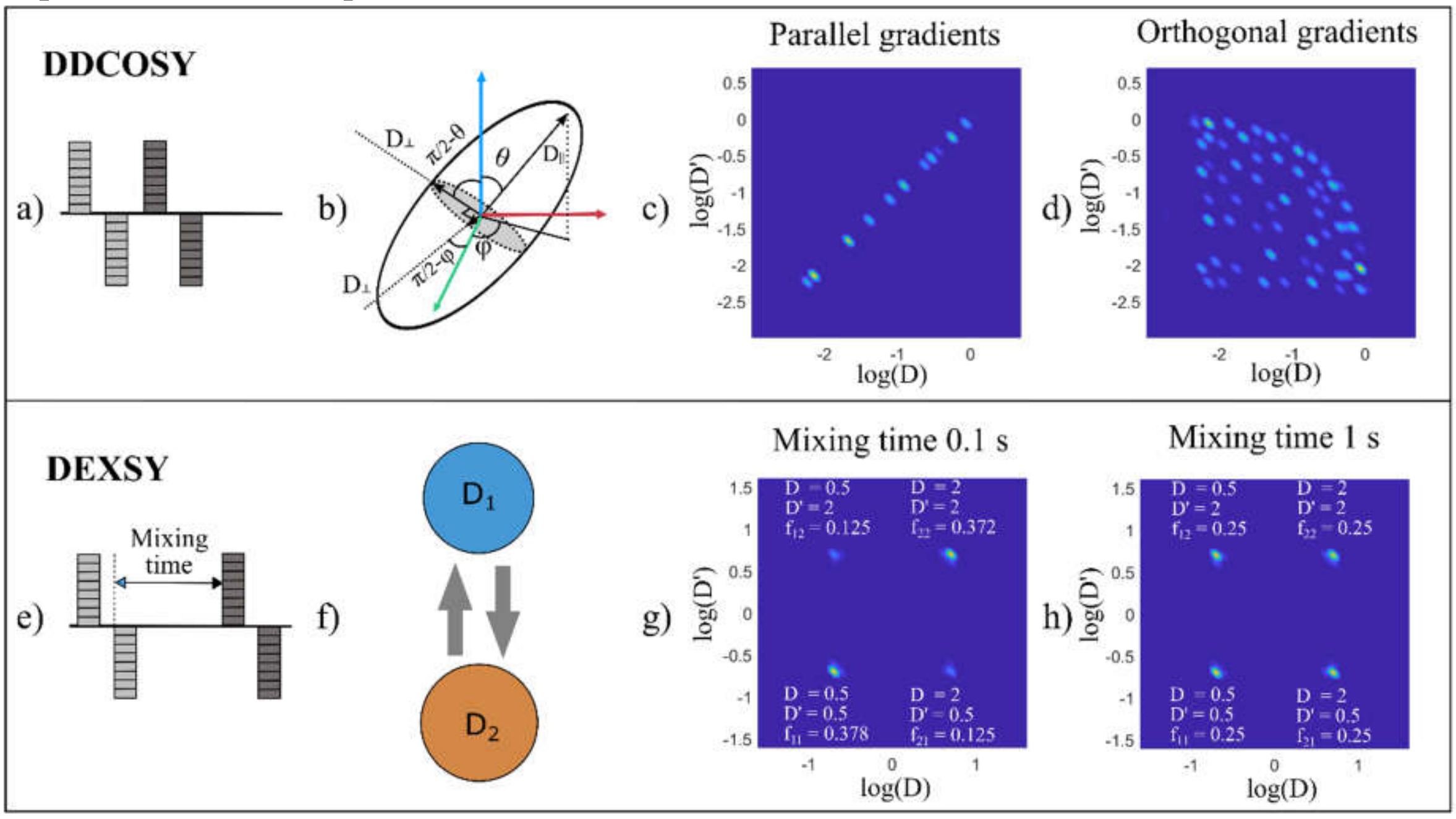

*Figure 6 a) Schematic representation of the DDE sequence used in the DDCOSY experiment. b) schematic representation of an anisotropic microdomain featuring Gaussian diffusion with DDE encoding along z and x direction. c)-d) 2D diffusion spectra obtained from a simulated DDCOSY experiment with parallel and orthogonal gradients for a substrate consisting of randomly oriented microdomains featuring Gaussian diffusion with $D_{\parallel}$ = 1 μm²/ms and D = 0.1 μm²/ms. The b-value of each gradient pair was varied from 0 to 5000 s/mm² in 100 steps. For this simulation the timing parameters of the sequence are not relevant given the diffusion is Gaussian. e) Schematic representation of the DDE sequence used in the DEXSY experiment. f) Schematic representation of two exchanging water pools. g)-h) 2D diffusion spectra obtained from simulated DEXSY experiments for a substrate consisting of two exchanging isotropic water pools ($D_1$ = 2 μm²/ms, $D_2$ = 0.5 μm²/ms, equilibrium volume fractions $p_1^{\infty} = p_2^{\infty}$ = 0.5 and exchange rates $k_{12} = k_{21}$ = 3.33 $s^{-1}$) at mixing times of 0.1s and 1s. The simulations were based on the Kärger model (Karger, Pfeifer et al. 1988) with the following sequence parameters: δ = 1ms, Δ = 5 ms, 40 b-value steps between 0 and 4000 s/mm² and two mixing times of 0.1s and 1s. All Inverse Laplace Transforms were performed using the MERA toolbox (https://github.com/markdoes/MERA).*

## 6.2 Diffusion Exchange Spectroscopy (DEXSY)

The DEXSY approach (Callaghan and Furo 2004), which aims to probe diffusion exchange, employs DDE sequences with parallel gradients in which the amplitude of each gradient pair is stepped independently, to compare the diffusion coefficients after a mixing interval $\tau_m$. The mixing time is usually chosen long enough to allow for exchange. Considering a toy model of two isotropic Gaussian water pools with diffusivities $D_1$ and $D_2$ as illustrated in Figure 6, the DDE signal in the absence of exchange can be written as:

$$S = f_1 e^{-(b_1+b_2)D_1} + f_2 e^{-(b_1+b_2)D_2}, \qquad (28)$$

where $f_1 + f_2 = 1$ are the volume fractions of the two compartments and are time independent. In the presence of exchange, assuming it occurs only during the mixing time, i.e. $\Delta << 1/k$ with $k$ being the exchange rate, the signal can be written as:

$$S = f_{1,1} e^{-(b_1+b_2)D_1} + f_{2,2} e^{-(b_1+b_2)D_2} + f_{1,2} e^{-b_1D_1-b_2D_2} + f_{2,1} e^{-b_1D_2-b_2D_1}, \qquad (29)$$

where $f_{1,1}$ and $f_{2,2}$ are the fraction of spins that resided only in compartment 1 and 2, respectively, while $f_{1,2}$ and $f_{2,1}$ are the fractions of spins which resided in compartment 1 during the first gradient pulse and in compartment 2 during the second one, and vice versa. The fractions depend on the mixing time and add to 1.

Thus, if there is no exchange between microdomains, the 2D diffusion spectrum obtained from an ILT of the DDE measurements will exhibit only diagonal peaks. If there is exchange, either between microdomains with different diffusivities as described in Eq. (29) or with different orientations, as detailed in (Callaghan 2011), then the 2D spectrum will also have off-diagonal peaks, which depend on the mixing time as illustrated in Figure 6. Thus, DEXSY can inform on exchange without making any direct modelling assumptions. A full description of exchanging water pools is beyond the scope of this work and can be found elsewhere (Karger, Pfeifer et al. 1988, Nilsson, Alerstam et al. 2010, Ning, Nilsson et al. 2018).

### 6.2.1 Preclinical applications

The first applications of DEXSY have been in material science, for example to study lyotropic liquid crystals (Callaghan and Furo 2004) or dextran exchange through polyelectrolyte multilayer capsules (Qiao, Galvosas et al. 2005, Galvosas, Qiao et al. 2007). Using a microcapillary phantom, Benjamini and Basser have shown that the MADCO framework with an acquisition consisting of 1D measurements followed by only four points in the 2D space can be successfully employed to map water exchange at the interface between microcapillaries and bulk water. The subsequent study by Cai et al reparametrized Eq. (29) in terms of the sum and difference of $b_1$ and $b_2$ to facilitate the estimation of the exchanging fractions, which can be estimated from only 4 data points in the 2D space after the diffusivities of the two water-pools have been measured from a 1D experiment (Cai, Benjamini et al. 2018).

DEXSY experiments have been employed only recently to study water microenvironments and exchange in biological tissue. For instance, Breen-Norris et al has derived a normalized diffusion exchange index (computed as the total area of exchange peaks divided by the total area of non-exchange peaks) from DEXSY measurements and employed it to study cell membrane permeability and exchange in yeast cells as well as in-vivo in a xenograft tumour model (Breen-Norris, Siow et al. 2020). Another recent study by Williamson el al employed DEXSY measurements in a constant gradient field to study microstructure and membrane permeability of live and fixed excised neonatal mouse spinal cords (Williamson, Ravin et al. 2019). The study included both full 2D measurements as well as the fast estimation of exchange fractions (Cai, Benjamini et al. 2018), to study the exchange properties at different mixing times. Fitting the Apparent Exchange Rate (AXR) model to the data (Eq. (30)) resulted in AXR values on the order of 100 $s^{-1}$. Moreover, measuring diffusion and exchange properties during tissue de-lipidation with Triton X showed that restricted diffusion is mainly due to lipid membranes.

## 6.3 Filter Exchange Spectroscopy / Imaging

A simplified version of the DEXSY experiment proposed by Åslund et al is the Filter Exchange SpectroscopY (FEXSY with its imaging counterpart FEXI) experiment using the same sequence layout but with the first filter gradient encoding set to a fixed amplitude (Aslund, Nowacka et al. 2009). The motivation for FEXSY is to reduce the multidimensional fitting problem for quantification of exchange in simple two component systems with different diffusivities suitable for in vivo imaging settings. In this framework, the first encoding is considered as a filter mainly reducing the signal from a fast component. The subsequent diffusion encoding probes the diffusivity of the remaining signal. A mixing time dependent increase in the diffusivity is expected if spins in the remaining slow component exchange into the fast component, as illustrated in Figure 7b. Assuming that each signal component exhibits mono-exponential behaviour, a two component gaussian model can be characterized by their signal fractions and diffusivities as well as the exchange time constant. Simplified further by Lasic et al, the Apparent Exchange Rate (AXR) model fitted to FEXSY data can be used to detect the time dependent level of mixing from the increase in diffusivity of the filtered signal in the low b range (Lasic, Nilsson et al. 2011, Scher, Reuveni et al. 2020). Thus, the apparent diffusivity after the filter has been applied ($\mathrm{ADC}'$) has the following mixing time dependence:

$$ADC'(t_m) = ADC(1-\sigma\cdot\exp(-AXR\cdot t_m)), \qquad (30)$$

where $\mathrm{ADC}'$ is the equilibrium apparent diffusivity either measured without a filter or after a sufficient mixing time $t_{\mathrm{m}}$. The *filter efficiency* σ reflects how clearly a fast diffusion coefficient is separated from the slow one after the filter.

### 6.3.1 Preclinical and clinical applications

First experiments were applied to yeast phantoms (Aslund, Nowacka et al. 2009, Lasic, Nilsson et al. 2011) and have been later implemented to study exchange properties of tissue in breast cancer tumours (Lasic, Oredsson et al. 2016) as well as in healthy brain and intracranial tumours (Nilsson, Lätt et al. 2013, Lampinen, Szczepankiewicz et al. 2017). For instance, Figure 7b illustrates the AXR and σ maps in the brain for a healthy volunteer (Nilsson, Lätt et al. 2013). Preclinical studies have also demonstrated the use of AXR as an MRI reporter of gene expression by adding genes for permeability modulating urea transporter to viral vectors injected into live animals (Schilling, Ros et al. 2017). Another recent study of FEXSY data in yeast cells and ex-vivo porcine optic nerve has shown that exchange rates mapped by varying the mixing time at a constant q value yields similar results to fitting the 2D measurement space obtained by varying both the mixing time and the q value of the second gradient pair, paving the way for fast acquisitions (Scher, Reuveni et al. 2020). A recent study from Bai et al investigated the ability of FEXI to map different types of diffusion exchange processed in the human brain by varying the strength of the filter gradient (Bai, Li et al. 2020). Specifically, a filter with a low b-value (250 $s/mm^2$) was used to map the exchange between the perfusion and diffusion water pools, while a higher filter with a b value of 900 $s/mm^2$, was used to map the exchange between the fast and slow tissue water pools. Indeed, the two types of filter provided different AXR and σ estimates, with the low filter analysis providing a high correlation between σ and the IVIM perfusion fraction.

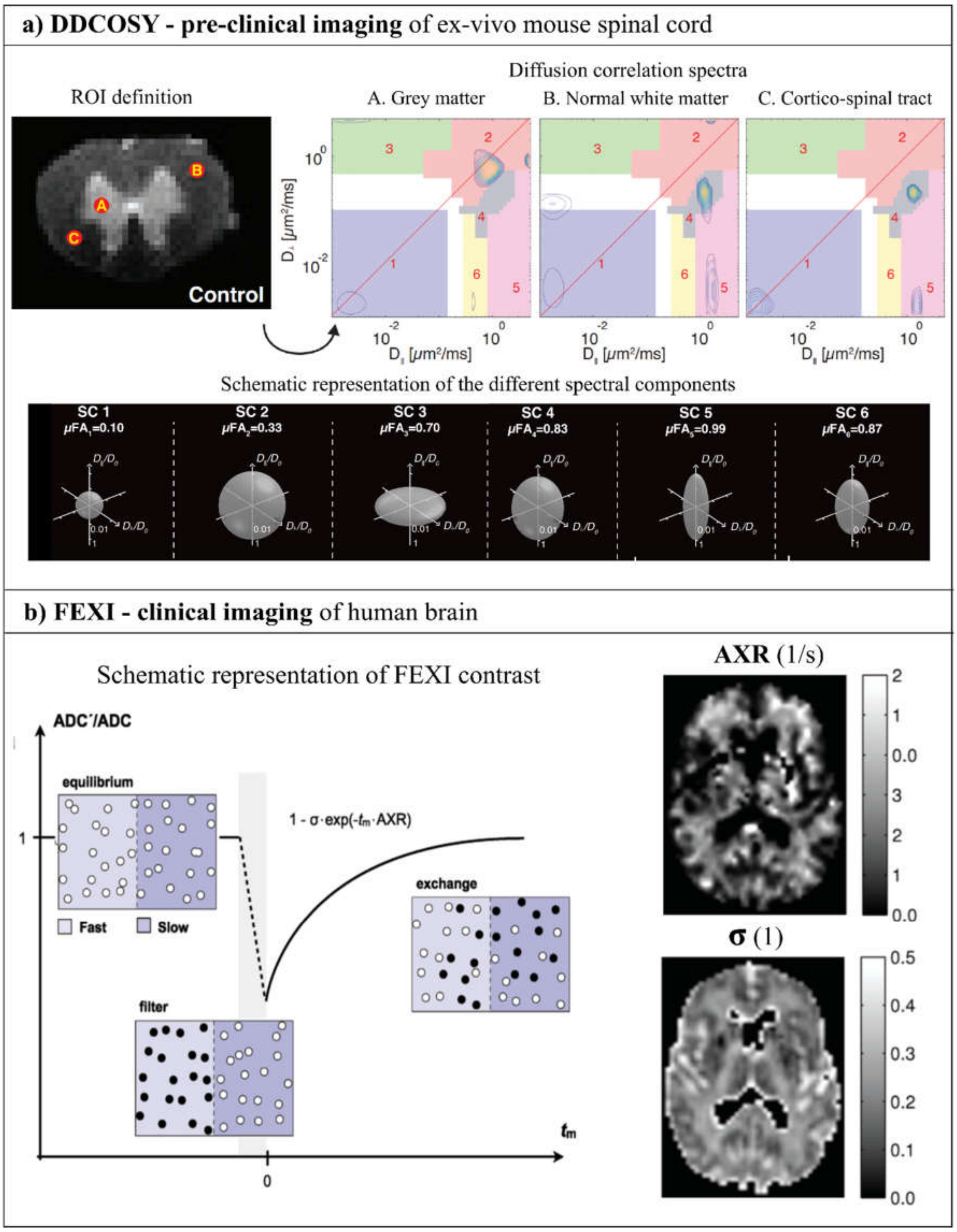

*Figure 7 a) Example of 2D diffusion-diffusion correlation maps derived from DDCOSY measurements in three ROIs of the neonatal mouse spinal cord, ex-vivo. The 2D probability distributions can be described by six distinct spectral components in the ($D_1$, $D_2$) space. Figure adapted from (Williamson, Ravin et al. 2019) b) Schematic representation of the FEXI contrast based on two exchanging water pools and in-vivo maps of AXR and filter efficiency (σ) in a healthy volunteer. Figure adapted from (Nilsson, Lätt et al. 2013).*

Some considerations for the experimental implementation of FEXSY/FEXI have been suggested. With a limited number of sampling points, the protocol may be sensitive to parameter settings, and the experimental design should be well related to the tissue or pathology of interest (Lampinen, Szczepankiewicz et al. 2017). Other time-dependent diffusion effects may also introduce additional modulation of the signal as possible interaction between displacements in the two encodings at short mixing times (utilized in pore size detection discussed in section 3.5). One proposed solution to this potential bias is the use of both a parallel and an anti-parallel filter to determine the size of the effect. While the first approaches of FEXSY/FEXI considered systems that where isotropic, underlying anisotropy may also affect the filtering effect and fitted AXR values which could be circumvented by considering measurements from uniformly distributed gradient directions. FEXSY/FEXI in general require long mixing times in the order of seconds and most implementations thus utilize STEAM sequences with longitudinal storage of the signal during the mixing time. Crusher gradients, necessary for maintaining the original signal pathway, may here introduce additional diffusion weighting counteracting the exchange driven increase in diffusivity. This effect, which may also bias other STEAM based DDE measurements, can be particularly pronounced in high resolution settings where additional analysis can be used to isolate and correct biases (Lasic, Lundell et al. 2018). In the STEAM based FEXI, T1 relaxation can also bias the ADC time dependence and consequently the estimated AXR. Some strategies, such as the recently proposed use of twice-refocused STEAM, may be employed to mitigate this bias (Martin, Endt et al. 2020)

## 6.4 Filter-Probe double diffusion encoding

Besides mapping exchange, DDE acquisitions where the first gradient pair acts as a filter have also been used to null the contribution of fast diffusing spins and reduce partial volume effects. For example, recent studies have proposed the use of orthogonal DDE experiments to resolve the axial diffusivity of coherent spinal cord microstructures from confounding partial volume effects from oedema and cerebrospinal fluid (Skinner, Kurpad et al. 2015, Budde, Skinner et al. 2017, Skinner, Kurpad et al. 2017). To achieve this effect, the first DDE gradient is applied perpendicular to the spinal cord supressing the signal from fast water diffusion processes, then the second DDE gradient is applied along the spinal cord to quantify the axonal diffusivity from non-suppressed water spins. This diffusion anisotropy measures were shown to highly correlate with histological-measured axonal injury (Skinner, Lee et al. 2018).

## 6.5 Velocity Exchange Spectroscopy (VEXSY)

The VEXSY experiment (Callaghan and Manz 1994, Manz, Seymour et al. 1997, Blumich, Callaghan et al. 2001) employs the same acquisition as DEXSY, but instead of analysing the data using a 2D ILT it performs a Fourier Transform in order to obtain the 2D displacement spectrum which can inform on the change in velocity from one diffusion period to the other. Ahlgren et al used a simplified VEXSY experiment on a clinical scanner to investigate velocity correlations in blood microcirculation causing signal attenuation (also known as the intravoxel incoherent motion (IVIM) effect) in the human brain (Ahlgren, Knutsson et al. 2016). They showed that the effect was removed in the anti-parallel condition and concluded that acceleration and higher order motion terms were negligible over the timescale of their measurement. A more detailed analysis of multidimensional (diffusion) NMR experiments is presented elsewhere, for example in (Callaghan 2011, Topgaard 2017).

## 6.6 Diffusion correlation via multidimensional diffusion encoding

Multi-dimensional diffusion correlation information can also be obtained with other types of diffusion encoding. For instance, triple diffusion encoding (TDE) which extends the DDE approach to three diffusion gradient pairs (initially proposed to achieve weighting by the trace of the diffusion tensor in one scan when the three gradient directions have orthogonal orientations (Mori and van Zijl 1995)), has been recently employed to study diffusion-diffusion correlation properties (Topgaard 2017). Specifically, de Almeida Martins and Topgaard used TDE yielding linear and isotropic encoding to map the joint probability distribution of the $D_{zz}$ diffusion tensor component and the isotropic diffusivity via ILT (de Almeida Martins and Topgaard 2016). Employing a change of variables, the authors used the same acquisition to compute the joint distribution of microdomain parallel and perpendicular diffusivities. Subsequent studies have extended the diffusion-diffusion correlation technique to include the angular distribution as well as relaxation parameters (de Almeida Martins and Topgaard 2018) and investigated different numerical inversion approaches (Reymbaut, Mezzani et al. 2020). Under the GPD, spectral modulation of gradient trajectories can be used to isolate time dependent diffusion from anisotropy (Scharff Nielsen, Dyrby et al. 2018, Lundell, Nilsson et al. 2019). An in-depth analysis of multi-dimensional diffusion encoding focused on different shapes of the b-tensor is presented in a separate review chapter.

# 7. DDE in metabolites spectroscopy

Most DDE experiments to date have been performed on water signals. While providing high sensitivity, these signals provide little specificity, as water is present in all cells and subcellular compartments, as well as in extra-cellular space and blood vessels. Below, we highlight how specificity towards microstructure can be enhanced using DDE-based magnetic resonance spectroscopy.

## 7.1 Metabolites diffusion-weighted magnetic resonance spectroscopy (dMRS)

*Metabolites* are the small molecule substrates, intermediates and products of *metabolism*: the set of life-sustaining chemical reactions in organisms. The chemical reactions of metabolism are organized into metabolic pathways, in which one chemical is transformed through a series of steps into another chemical, for three main purpose: the conversion of food to energy to run cellular processes; the conversion of food/fuel to building blocks for proteins, lipids, nucleic acids, and some carbohydrates; and the elimination of nitrogenous wastes (Siesjo 1979). Metabolites are commonly divided into either endogenous (produced by the host organism), or exogenous (with metabolites of foreign substances such as drugs termed xenometabolites). Within the magnetic resonance context, recent improvements in scanner hardware has allowed the implementation of magnetic resonance spectroscopy (MRS) at very high field in vivo, yielding highly-resolved spectra for the accurate quantification of *brain metabolites* (Govindaraju, Young et al. 2000, Tkác, Oz et al. 2009, Duarte, Lei et al. 2012). MRS therefore can be used to directly, specifically and non-invasively access the brain biochemistry (Ronen and Valette 2015, Palombo, Shemesh et al. 2018).

In MRI, frequency encoding is used to determine the spatial location of spins. In MRS, the frequency domain is used to read out chemical signatures of the MR-responsive nuclei present in the tissue. (Ronen and Valette 2015, Cao and Wu 2017). Thus, rather than images, MRS data are usually presented as line spectra (Figure 8), the area under each peak representing the relative concentration of nuclei detected for a given chemical species within the selected MRS voxel (usually, the volume of MRS voxels is ~10-100 times larger than typical MRI voxels of 1x1x1 $mm^3$ in clinical systems). Protons are more commonly used for spectroscopy because of their high natural abundance in organic structures and their high magnetic sensitivity when compared with other nuclei, such as phosphorus, sodium or others, which require specialized coils and amplifiers for observation (De Graaf 2007).

Like its water dMRI counterpart, it is possible to impart sensitivity towards metabolite diffusion, for instance by using SDE techniques together with MRS methods. Such sequences can be generally referred to as diffusion-weighted MRS (dMRS). Examples of some spectra acquired at different diffusion weightings are reported in (Figure 8). From the estimation of the area under each peak, it is possible to evaluate the signal attenuation due to diffusion for a specific metabolite which corresponds to the assigned peak. In this way it is possible to obtain, for each metabolite, or group of metabolites, a normalized diffusion-weighted signal attenuation, S(b)/S(b=0), as a function of b (Figure 8) (or q-value, in an analogue of QSI (Assaf and Cohen 1998)) completely equivalent to the typical diffusion-weighted signal attenuation curves that are analysed in dMRI to reconstruct parametric maps of brain microstructure features.

This is of particular interest because, unlike water, which is present in all relevant compartments, some of the endogenous brain metabolites are *purely intracellular* and/or preferentially confined in *specific cell compartments*, as in the case of N-acetylaspartate (NAA) and glutamate (Glu), which are prominent neuronal markers, or myo-inositol (Ins) and choline compounds (tCho), which are more associated with glial cells (Choi, Dedeoglu et al. 2007) (Figure 8). Although involved in metabolism, indirect measurements (Moonen, Van Zijl et al. 1990, Marchadour, Brouillet et al. 2012, Ronen, Ercan et al. 2013, Branzoli, Ercan et al. 2014, Najac, Marchadour et al. 2014 , Ronen, Budde et al. 2014, Ercan, Magro-Checa et al. 2016, Ianus 2016, Palombo, Ligneul et al. 2016, Ligneul and Valette 2017, Valette, Ligneul et al. 2018) have shown that metabolites are constantly exploring their local environment under the effect of diffusion, thus potentially reporting precious information on several features of cell microstructure such as cell fibre length (Palombo, Ligneul et al. 2016) or diameter (Palombo, Ligneul et al. 2017). This enables probing the microstructure of specific cell types, which might be highly relevant in a neuropathological context where severe morphological alterations of specific cells can occur, for instance neuronal atrophy or neuroinflammation (Ercan, Magro-Checa et al. 2016, Ligneul, Palombo et al. 2019).

Different SDE dMRS approaches have been proposed to quantify cell microstructure, such as diffusion tensor spectroscopy (Ellegood, Hanstock et al. 2005, Ellegood, Hanstock et al. 2006, Ellegood, McKay et al. 2007, Hanstock and Beaulieu 2020) (mainly to quantify the degree of macroscopic anisotropy of brain metabolites diffusion), bi-exponential diffusion modelling(Assaf and Cohen 1998, Assaf and Cohen 1998), high-b value experiments (mainly to probe fibre diameter but also presumably yielding some sensitivity to cell body diameter (Palombo, Ligneul et al. 2017, Ligneul, Palombo et al. 2019, Lundell, Ingo et al. 2020, Palombo, Ianus et al. 2020), or measurements of the apparent diffusion coefficient (ADC) up to very long diffusion times to probe long-range cell structure (potentially enabling quantification of cell fibre branching and length (Najac, Branzoli et al. 2016, Palombo, Ligneul et al. 2016, Ligneul, Palombo et al. 2019). Such approaches were recently used to quantitatively estimate alterations of astrocytic morphology in a mouse model of reactive astrocytes, where especially the diffusion of Ins was significantly different compared to the control group (Ligneul, Palombo et al. 2019). Furthermore, dMRS experiments employing oscillating gradients have been used to measure metabolites ADC at very short time-scales (~0.5–1 ms) and access information about cytoplasm viscosity (Valette, Ligneul et al. 2018), in the rat (Marchadour, Brouillet et al. 2012) and mouse brain(Ligneul and Valette 2017, Doring and Kreis 2019). These studies showed that metabolites ADC increased by ~50% when the frequency increased from ~20 to ~250 Hz for NAA, tCho, and Creatine, (in the mouse brain, also for Ins and Taurine), approaching diffusivity values of ~0.2–0.30 $\mu m^2/ms$ at the highest frequency. Modelling these data by using frequency-domain formalism for diffusion in cylinders or spherical pores (Stepisnik 1981, Stepisnik 1993, Callaghan and Stepisnik 1995), the authors estimated typical asymptotic intracellular diffusivities to be ~0.5–0.6 $\mu m^2$/ms, i.e., corresponding to a low-viscosity cytosol, less than twice the viscosity of pure water. Similar diffusivity values in pure WM have also been obtained from data measured in the long diffusion time regime (Ronen, Ercan et al. 2013, Lundell, Ingo et al. 2020). This is in good agreement with fluorescence-based estimates of fluid-phase cytoplasm viscosity being quite similar to bulk water (Fushimi and Verkman 1991, Luby-Phelps, Mujumdar et al. 1993). These results were also recently confirmed by studies in human brain (even though at lower frequencies) (Doring and Kreis 2019).

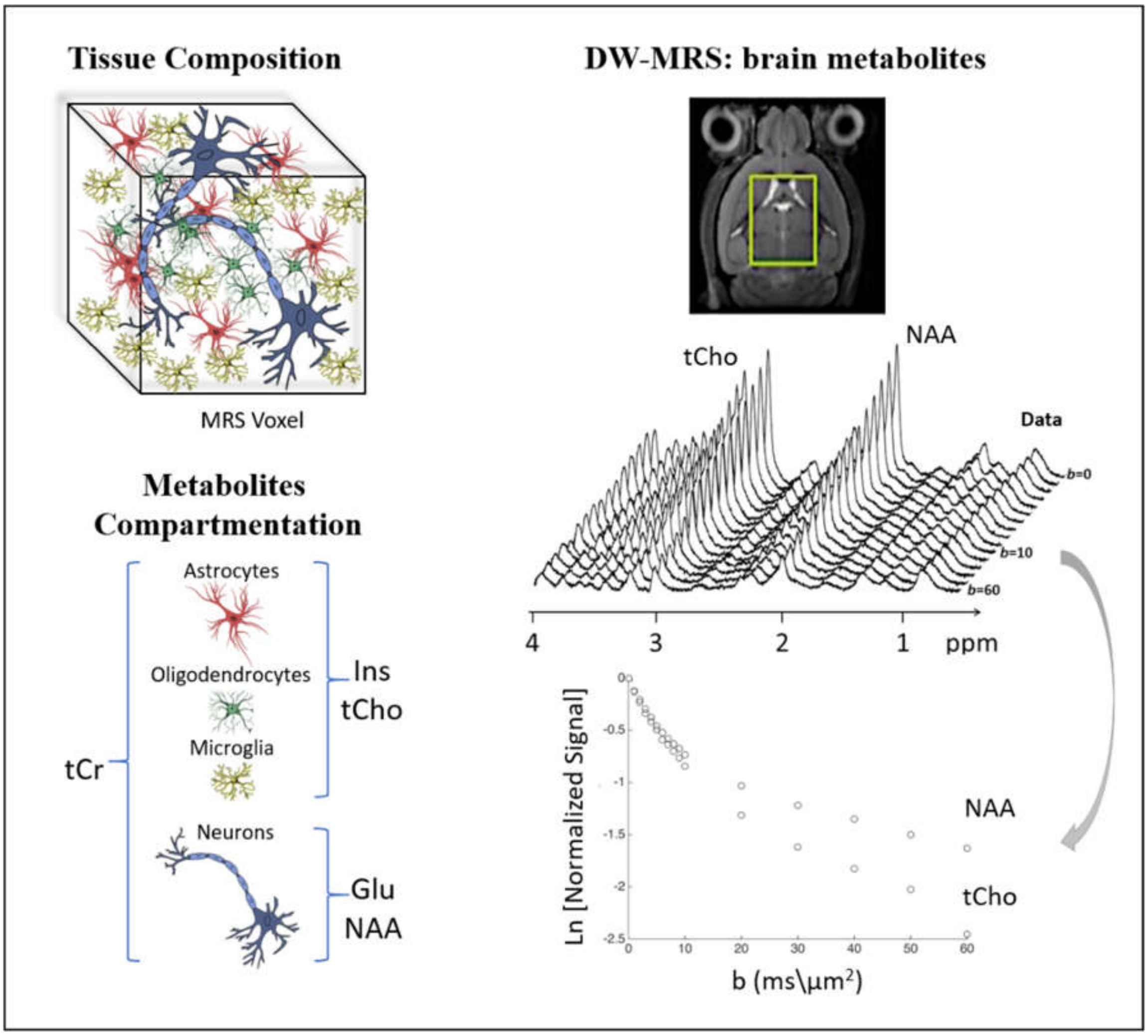

*Figure 8 DW-MRS provides diffusion-weighted spectra within a large voxel (typically some tens of mL) with very heterogeneous tissue composition. Each peak in the spectra corresponds to specific metabolites. The x-axis denotes the frequency shift localizing the metabolite in parts per million, ppm, while the vertical y-axis plots the relative signal amplitude or concentrations for the various metabolites, i.e., the height of the peak reflects the amount of the metabolite. Integrating the area under each peak a diffusion-weighted signal as a function of diffusion-weighting b can be measured for each metabolite. Given the unique cell type specific compartmentation of some metabolites, DW-MRS can provide higher specificity to intra-cellular tissue properties than conventional water based DW-MRI.*

Although insightful and powerful techniques to quantify cell-type specific morphological alterations, SDE based measurements often lack the ability to reflect microscopic anisotropies, especially when considering the signal originating from big spectroscopic voxels comprised of highly heterogeneous microstructure. In contrast, DDE methods are emerging as promising techniques for decoupling the microscopic anisotropy from the overall orientation dispersion of the studied tissue (see previous **Sections 3** and **4**) and have been recently extended to dMRS applications. An overview of the commonly used DDE based dMRS sequences is reported in the next section.

## 7.2 Overview of sequences for DDE based metabolites dMRS

Like for dMRI, DDE based dMRS measurements can be obtained by modifying SDE dMRS sequences to accommodate for the two diffusion blocks. In particular, Shemesh et al. (Shemesh, Rosenberg et al. 2014) proposed an efficient modular preparation-contrast-localization-acquisition layout for dMRS sequences that easily enables the implementation of DDE filters in the contrast module, keeping the preparation and localization blocks substantially unchanged (Figure 9a). Note that this design also mitigates cross-terms arising from internal gradients and leads to spatially localized spectra, reflecting the diffusion contrast imposed with a very high sensitivity (SNR>50:1 for the NAA peak, at 21.1T, with 6 seconds acquisition – 8 scans, 5x5x5 $mm^3$ voxel).

In a following work, Shemesh et al. (Shemesh, Rosenberg et al. 2017) used the same sequence design but with the variant involving a Carr-Purcell-Meiboom-Gill (CPMG) train during the DDE filter to selectively excite and monitor the diffusion process of NAA and Ins only. The CPMG module was introduced to mitigate potential interactions between the DDE gradients and susceptibility-induced gradients, i.e. internal gradients that at ultrahigh fields (e.g. 21.1 T) could not be neglected and could distort the DDE curves if unaddressed.

More recently, Vincent et al. (Vincent, Palombo et al. 2020) advanced the DDE MRS methodology presented in (Shemesh, Rosenberg et al. 2014), by employing a variable power with optimized relaxation delays (VAPOR) module for water suppression and conventional radio frequency pulses to detect more metabolites (reliable measurements from tNAA, tCr, tCho, Lac, Ins and Tau). Moreover, a more refined state-of-the-art post-processing pipeline was used for accurate diffusion-weighted signal quantification including, specifically, individual scan phase correction, LCModel analysis and macromolecule (MM) signal quantification, openly available at https://github.com/Meli64/DDE-data-processing.

The works from Shemesh et al. (Shemesh, Rosenberg et al. 2014, Shemesh, Rosenberg et al. 2017) and Vincent et al. (Vincent, Palombo et al. 2020) are examples of DDE based dMRS sequences optimized for metabolites diffusion measurements on preclinical scanners and for preclinical applications. Translating these approaches to clinical scanners and applications is challenging. For example, the optimal localization achieved by using the LASER modulus cannot be employed in clinical scans because the specific absorption rate (SAR) of the fully adiabatic radio frequency pulses is too high (Ronen and Valette 2015, Cao and Wu 2017). A compromise has been recently proposed by Lundell et al. (Lundell, Webb et al. 2018) and Najac et al. (Najac, Lundell et al. 2019) who proposed a DDE sequence based on a partially adiabatic semi-LASER sequence (Figure 9b). In this implementation, the diffusion-weighted blocks are based on bipolar gradients and inserted in a preparation module with four 180° pulses. The gradient direction of the first block is usually fixed, e.g. along the x axis, while the gradient direction of the second block revolves in N (e.g. 12) angular steps on a circle that includes the first gradient direction. This design enables efficient localization, but it is not immune to cross-term bias. To compensate for such cross-terms between diffusion and imaging/background gradients, data were acquired with positive and negative diffusion gradients and the geometric mean of the two signals was used for the analysis.

## 7.3 Applications in preclinical setting.

Because it is still a relatively novel method, there are only a few preclinical applications of DDE MRS of brain metabolites. A first application was pioneered by Shemesh et al. (Shemesh, Rosenberg et al. 2014) to non-invasively follow microstructural alterations of ischemic tissues in a stroke animal model, using the sequence in Figure 9a. Signal modulation as a function of φ (i.e. the relative angle between the gradient orientations) unambiguously demonstrated that metabolites diffuse in elongated compartments and that the microscopic anisotropy for NAA, tCho and total Creatine (tCr) dramatically increases 24 h after the onset of ischemia (Figure 9a). Moreover, the results revealed that Lac diffuses in a different environment than the remaining metabolites. The authors speculate that this could help elucidate the nature of conventional diffusion experiments reporting that upon ischaemia water exhibits a decreased diffusivity; in the metabolite-based measurements, this decrease appears to be correlated to increases in the apparent eccentricity of the underlying microstructures.

A subsequent study by Shemesh et al. (Shemesh, Rosenberg et al. 2017) investigated the potential of DDE MRS to quantitatively extract cell fibre diameter (see Section 3) in vivo in the healthy mouse brain. The authors used the sequence in Figure 9a for measuring the diffusion properties of NAA and Ins to allow distinguishing neuronal (NAA) and astrocytic (Ins) compartments. By analysing the DDE signal modulation using a simple geometrical model of randomly oriented infinite cylinders, the estimated cell-fibre diameter was smaller for NAA (~0.4 μm) than for Ins (~3.1 μm), which was well in line with diameter values inferred from SDE high-b measurements in the study by Palombo et al. (Palombo, Ligneul et al. 2017): 1.24±0.24 μm for tNAA and 3.11±0.16 μm for Ins.

One drawback of Shemesh et al.'s results was a high uncertainty in the estimated cell-fibre diameters, suggesting the necessity of better measurements and/or improved modelling for more reliably estimates. Driven by this motivation, Vincent et al. (Vincent, Palombo et al. 2020) used the improved DDE MRS methodology measurements to revisit Shemesh et al.'s results in vivo in healthy mouse brain and measure signal modulations for other metabolites of interest. By using a sequence similar to the one in Figure 9a and a state-of-the-art processing and analysis pipeline (including individual spectra rephasing, eddy current correction and LCmodel quantification), Vincent et al. showed that additional features of the cell microstructure, such as cell body diameter, fibre length and branching may also influence the DDE signal. Specifically, by harnessing advanced 3D cell models (Palombo, Alexander et al. 2019), the authors pinpointed which features of cell morphology may influence the most the angular dependence of the DDE signal. Their results showed that while the infinite cylinder model poorly fits the experimental data, incorporating branched fibre structure in the model allows more realistic interpretation of the DDE signal. Interestingly, data acquired in the short mixing time regime suggested that some sensitivity to cell body diameter might be retrieved, in agreement with some recent findings using water based dMRI (Palombo, Ianus et al. 2020). A recent simulation study that employed similar 3D cell models to systematically investigate the impact of soma size and branching of cellular projections on the DDE signal, also corroborates these results and showed a measurable effect of branching order at long mixing time and of soma size at short mixing time (Ianus, Alexander et al. 2020).

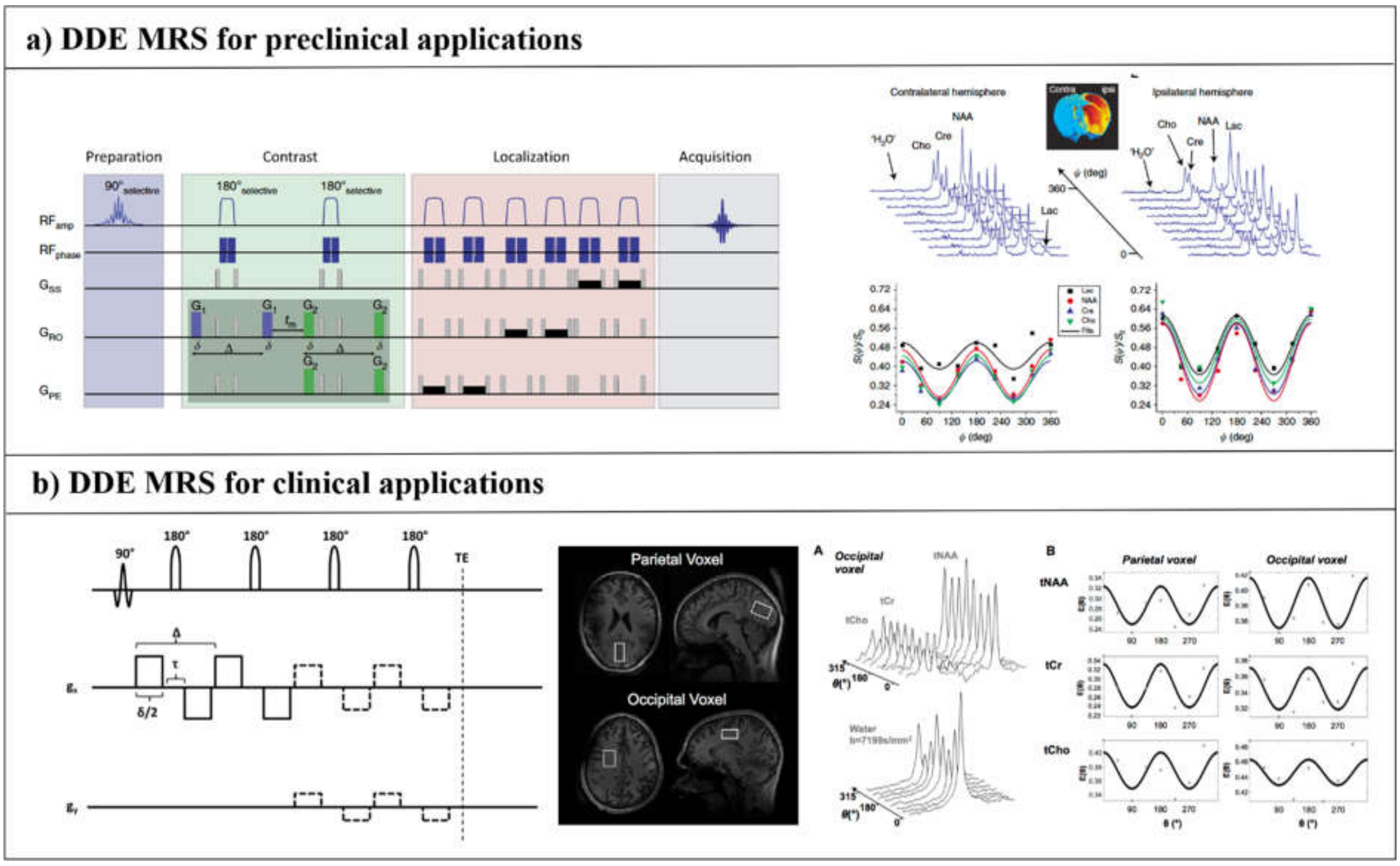

*Figure 9 a) Diagram of the sequence design proposed in (Shemesh, Rosenberg et al. 2014), where the preparation, DDE weighting (contrast), localization and acquisition are separated in blocks. The separation of the DDE weighting and the localization block allows the mitigation of the cross-terms. The angular modulation of the DDE signals obtained in a preclinical experiment in the rat brain where stroke was induced are also shown for four major metabolites (adapted from (Shemesh, Rosenberg et al. 2014)). b) For clinical translation of DDE dMRS the sequence in a) cannot be used due to too high SAR and other limitations. Therefore, a simpler sequence has been used in (Lundell, Webb et al. 2018) to obtain the angular modulation of three main metabolites (tNAA, tCr and tCho) in the healthy human brain (adapted from (Najac, Lundell et al. 2019)).*

## 7.4 Applications in clinical setting.

Like for preclinical applications, there are currently only a few preliminary works developing metabolites DDE MRS measurements for in vivo clinical applications.

A first very interesting investigation of metabolites diffusion microscopic anisotropy in vivo in the healthy human brain has been recently reported by Lundell et al. (Lundell, Webb et al. 2018). By using the DDE MRS sequence in Figure 9b, Lundell et al. quantified anisotropic intracellular diffusion in the human brain (5 healthy volunteers) at 7 T. Three metabolites were reliably measured: tNAA, tCr and phosphor-choline (PCho) in a large voxel positioned in the parietal region and containing mostly highly dispersed white matter tissue. The astrocytic PCho diffusivity showed somewhat lower microscopic anisotropy (0.50±0.45) compared to the neuronal tNAA signal (0.97±0.02), suggesting an additional contribution from isotropic cell bodies in white matter. Moreover, water microscopic anisotropy was close to the tNAA values (0.96±0.03) reflecting that fast extracellular components are filtered out at this relatively high b-value. In agreement with previous preclinical studies (see Section 7.3), Lundell et al.'s results support a main fibrous component (high microscopic anisotropy), although also suggesting a more complex geometry of astrocytes (Ingo, Brink et al. 2018) that could include isotropic compartments (such as cell bodies) or branching and undulating fibrous structures. Note that this work had some limitations, including unaccounted effects from residual macroscopic anisotropy, time dependent effects at short mixing times, and the use of only one non-zero b value (7192 s/mm$^2$).

To overcome some of the above limitations, Najac et al. (Najac, Lundell et al. 2019) proposed an extended version of this experiment to measure the microscopic anisotropy at different diffusion weighting in both white and grey matter. In their work, Najac et al. positioned a 9 mL voxel either in a white matter region within the parietal lobe (4 healthy volunteers) or within the occipital cortex (3 healthy volunteers). The two voxels were chosen to comprise two different tissue compositions: white matter / gray matter = ~80% / 20%; and ~ 50% / 45%, respectively (the residual volume fraction is occupied by cerebrospinal fluid) (Figure 9b). The same sequence and scanner as in (Lundell, Webb et al. 2018) were used, but with 4 non-zero b values in the water acquisition which is expected to reflect a more complex multi component scenario (918, 2066, 4050 and 7199s/mm$^2$). Also, total choline compounds (tCho) were this time quantified instead of only PCho as in (Lundell, Webb et al. 2018). The results showed no significant differences between metabolites microscopic anisotropy in both voxel locations, with average values of 0.85±0.10. Concerning water, at very high b (7199 s/mm$^2$), parietal water microscopic anisotropy was about 0.90, similar to metabolite microscopic anisotropy in both voxel locations, confirming that the extracellular compartment at very high b is essentially fully suppressed and the intracellular compartment exhibits microscopically anisotropic geometry. Moreover, in the occipital voxel, water microscopic anisotropy at the highest b value (0.715±0.080) was significantly lower compared to parietal voxel (0.918±0.007), suggesting that isotropic cell body contribution or intercompartmental exchange might play a role in GM, even at relatively short mixing/diffusion times. At low b values, water microscopic anisotropy was significantly decreased in both voxel locations (0.836±0.068 and 0.250±0.115, for parietal and occipital voxels respectively at b = 918 s/mm$^2$), indicating that the extracellular space is less microscopically anisotropic in both white and grey matter. At the lowest b, water microscopic anisotropy in the parietal voxel was 4x higher compared to the occipital voxel, showing that the extracellular space in white matter is highly anisotropic, or conversely, that the tortuosity in the direction perpendicular to the fibre propagation direction significantly affects water diffusion.

# 8. Practicalities (sequence preparation, readout, time constraints, etc)

As we have seen in previous sections, a number of phenomena related to diffusion and flow can potentially influence the DDE experiment. Many of the desired DDE signal changes are on the order of a couple of percent of the signal or even lower, which, for reliable detection, (and quantification), demand a careful experimental design. In realistic experimental conditions, the effect sizes should be contrasted with the limitations of the practical setting. A simple NMR experiment on phantoms or ex-vivo tissue with ample signal to noise will greatly differ from imaging experiments on animals or humans in vivo with different limitations arising, which we consider in detail below.

## 8.1 Contrast to noise

The success of DDE acquisitions can be thought of in terms of the fundamental contrast-to-noise (CNR) and the factors that affect it. As in any experiment, the CNR should be sufficiently high to reliably detect the effects being sought. In most DDE contexts, this usually will entail a difference of signals: for example, the difference of parallel and antiparallel gradient orientations to measure size, or the difference between parallel and perpendicular signals to measure microscopic anisotropy. Simply put, a sufficiently high CNR would involve the DDE signal difference being larger than the noise levels, e.g., $CNR_{DDE} = \frac{|S_A - S_B|}{\eta}$ where $\eta$ is the rician-biased noise level and $S_A$ and $S_B$ are DDE signals (in some cases these will be averaged signals, e.g., as in the 5-design). To illustrate this effect, Figure 10a presents the reproducibility analysis of μFA performed by Kerkela et al for DDE data acquired in an ex-vivo mouse brain (Kerkelä, Henriques et al. 2020). The results clearly show more uncertainty in the parameter estimates for smaller values of μFA, for which the signal difference $|S_A$ - $S_B|$ is close to zero. Several factors can be identified that increase the signal difference quite generally:

- The b/q-value: the higher it is, the higher the signal contrast (but also the significance of higher order terms). Note that in the short pulse approximation the q-value can only be increased by stronger gradients, while the b-value can be increased through either gradient strength and/or diffusion time. Outside this regime b/q-values can also be increased by increasing the diffusion gradient duration.
- Short diffusion gradient durations (δ) and long gradient separation Δ. When these are asymptotically "infinite" (short for δ and long for Δ), the signal difference is at its maximum for a given q-value;
- Extremities of mixing times, depending on the experiments (e.g., long mixing time for microscopic anisotropy, zero mixing time for measuring sizes), which can modulate the signal differences.

In all of the above, "short" and "long" are considered vis-à-vis the diffusivity and the "effective compartment size", e.g., $T \propto \frac{L_c^2}{nD_0}$ where T can be replaced by mixing time, diffusion times or diffusion gradient durations, $L_c$ is some general correlation length, n is proportional to the dimension and $D_0$ is the effective free diffusivity in the system. We note in passing that time-dependent effects can be of interest, as they add a dimension that may assist in portraying the microstructural features of the samples.

In addition, CNR can be enhanced by multiplexing (i.e., measuring more points on the signal difference curves) or signal averaging which will entail a SNR gain of $\sqrt{N}$ where N is the number of measurements. Nevertheless, when averaging magnitude data, one also needs to pay attention to the effect of noise floor which can bias the parameter estimates, or alternatively to use real and/or complex data instead (Eichner, Cauley et al. 2015, Fan, Nummenmaa et al. 2020). Another option is to perform the experiments at higher magnet field to increase the absolute signal to noise, although this also results in shorter relaxation times. All these can effectively reduce η and enhance CNR. The absolute SNR can be enhanced through minimizing TE (which is inherently quite difficult for a DDE) and maximizing TR, the latter typically at the expense of longer experimental durations.

## 8.2 CNR: preclinical scanner considerations

In the context of preclinical MRI, gradient amplitudes are typically higher (~300-600 mT/m on many horizontal bore scanners and >1500 mT/m on most vertical bore systems). In addition, preclinical systems are typically of higher field, e.g., >7 T. Therefore, a wider range of experimental parameters can be obtained, e.g., the gradient durations can be kept small to fulfil short gradient pulse approximations; this paves the way towards experiments that in imaging are much more demanding, for instance, measuring diffraction patterns or imaging pore density functions. However, the higher field experiments suffer from inherently shorter T2s, such that the effective TEs need to be kept shorter. Multiple stimulated-echo experiments have been performed to minimize TE while maximizing diffusion times and mixing time, mainly in spectroscopic mode due to the inherently lower signal to noise of stimulated echo acquisition strategies.

DDE MRI has mainly been performed in small animals in-vivo. It is important to ensure that physiological effects such as animal temperature are monitored and kept constant, due to the strong gradients applied. We also note that in our experience, respiratory gating is highly recommended to suppress bulk motion. Using multiple stimulated echoes is not very feasible in contemporary in-vivo imaging, so TE considerations must be kept in mind to enable sufficiently robust CNR to be recorded; typically, this would involve diffusion times of ~ < 20 ms and similarly mixing times of up to 20 ms, if using a spin echo preparation. With EPI readouts, the total TE of such a sequence would already exceed 60 ms, which, at 9.4 T, can significantly affect signal to noise and hence contrast to noise. If the sequences are implemented in a stimulated echo preparation, then the mixing times could be longer, but other issues need to be considered, as later discussed in Section 8.5. The relatively recent usage of cryogenic coils that afford significant signal to noise enhancements is an excellent (albeit expensive) solution to signal to noise dropouts (Baltes, Radzwill et al. 2009, Langhauser, Heiler et al. 2012, Nunes, Ianus et al. 2019). In addition, given that most DDE acquisitions in imaging contexts would require rotationally invariant schemes, the DDE sequence typically lends itself nicely towards denoising strategies as it requires at least tens of images (e.g. 72 directions for the 5 design plus several non-weighted images) which can assist in the reconstruction of its parametric maps.

## 8.3 CNR: Clinical scanner constraints

The clinical scanners pose two fundamental limitations towards acquiring robust DDE data. First, the gradient system needs to be able to actually deliver the amplitude and slew rate needed to achieve a measurable contrast. Since clinical gradient system amplitudes are quite limited as compared to their preclinical counterparts, this prerequisite nearly automatically entails long gradient pulse durations, on the order of several tens of milliseconds. The sample's relaxation properties then become crucial to take into account since the partial signal fraction carrying the effect of interest must provide sufficiently low noise levels to be measurable. Second, in clinical in vivo settings, scan times are nearly always an issue. The limited scanning time and possibly the gradient amplifier duty cycle (i.e. how long and strong gradients can be applied over time) will dictate the possible number of repetitions or parameter settings tested during a limited scan time. Standard clinical scanners operate at typically 40-80 mT/m with experimental systems capable of 200-300 mT/m (McNab, Edlow et al. 2013, Jones, Alexander et al. 2018, Weiger, Overweg et al. 2018, Foo, Tan et al. 2020). Peripheral nerve stimulations in the vicinity of large volume coverage gradient systems is a greater concern in humans and slew rates are therefore typically limited to the range of 200 T/m/s. Gradient coils with smaller volume coverage can be used to overcome this limitation where slew rates up to 1200 T/m/s has been achieved in vivo (Weiger, Overweg et al. 2018, Foo, Tan et al. 2020). Translating an animal DDE setup (limitations discussed below) to humans is therefore challenged by a roughly tenfold decrease in gradient performance.

The gradient field generally perform optimally in the isocentre but tapers off in the outskirts of the field of view. This may result in a diffusion weighting that potentially vary significantly over spatial locations or directions, the latter for instance violating necessary assumptions for a powder average. While the effect can be corrected for DTI analysis (Bammer, Auer et al. 2002, Holland, Kuperman et al. 2009), the effect becomes less traceable in higher order terms where at least the magnitude of the bias can be estimated from simulating signals with and without gradient nonlinearities (Mesri, David et al. 2020). A different effect also contributing to spatially varying gradient fields are concomitant fields, or Maxwell terms, which provide additional orthogonal gradient components away from the isocentre (Bernstein, Zhou et al. 1998, Baron, Lebel et al. 2012, Szczepankiewicz, Westin et al. 2019). One critical consequence of concomitant fields in diffusion encodings is the dependence on polarity which means that a positive gradient pulse will not be fully refocused by an equivalent negative pulse leading to a local signal dropout not related to diffusion. DDE experiments where the single diffusion encoding blocks are made with bipolar gradients may therefore be susceptible to this artifact, while an encoding with monopolar pulses around an inversion pulse compensate the effect. Consequently, any encoding waveform where the same amplitude, direction and duration is repeated at any time after inversion is by construction corrected. Note that gradient contributions from concomitant fields are generally small in terms of diffusion weighting. Nonetheless, local signal reductions from blurring the k-space may lead to substantial misinterpretation of signal attenuation unless appropriate corrections are done at the experimental stage (Baron, Lebel et al. 2012, Szczepankiewicz, Westin et al. 2019).

## 8.4 Signal to noise trade-offs

To ensure that DDE experiments provide robust metrics, the initial signal to noise ratio (SNR) must be as high as possible to enable a good CNR (as explained above). In most DDE contexts, the SNR involves intricate trade-offs between various experimental parameters. For instance, the diffusion times in a spin echo contribute to long relaxation times. Since the SNR is ultimately dictated in these sequences by the transversal relaxation time, T2, a balance needs to be struck between achieving sufficient diffusion weighting (long diffusion times) and the signal loss driven by the concomitant prolongation of TE. Practical echo times in clinical systems are generally for water in white matter below 150 ms in a human at 3T or even lower at higher fields or in post mortem samples where T2 is shorter. In preclinical systems, magnetic fields are typically >7T which renders T2s even shorter, typically ~30-60 ms. Therefore, shorter TEs are required, which can be compensated for by the stronger gradient amplitudes typically available on preclinical systems.

To mitigate TE effects, stimulated echo (STE) based sequences for DDE have been employed, mainly in spectroscopic mode. The STE benefits from dramatic shortening of the TE; however, given that the DDE sequence has two diffusion encodings and a mixing time, an efficient STE would require two (for short mixing time) or three (for long mixing time) STE epochs within the pulse sequence. Given that STE harnessing crusher gradients entail a $\left(\frac{1}{2}\right)^N$ signal decrease (with N being the number of STE epochs along a sequence), the sensitivity of such sequence can suffer dramatically. In spectroscopic mode, a loss of nearly an order of magnitude can still be acceptable and STE “templates” can be harnessed fruitfully; however, for imaging this is usually detrimental.

## 8.5 Bias

Once a sufficient CNR and SNR have been achieved, further considerations should be made to ensure that the measured contrast is not biased by other effects. Bias in DDE can be broadly categorized to sample-related bias (e.g. effects from internal gradients, flow, exchange and local diffusivity) and sequence-related bias (e.g. additional unaccounted sources of diffusion weightings).

**Background gradients**

The microstructure itself may impart internal gradients that can be significant, especially at ultrahigh fields. These can cause significant bias in the measured DDE signals, as internal gradient distributions are not necessarily isotropic. Hence, their interactions (cross terms) with the DDE gradient waveform may cause significant bias. For instance, Shemesh and Cohen (Shemesh and Cohen 2011) showed how the bell-shaped curves expected for simple DDE acquisitions at short $\tau_m$ can be completely inverted by internal gradients. Similarly, Lawrenz and Finsterbuch showed similar effects in human experiments, where background gradients can be particularly detrimental in gray matter close to CSF, bone and air interfaces (Lawrenz and Finsterbusch 2015), as illustrated in Figure 10b. When such effects are expected or noticed, pulse sequences that cancel such cross-terms need to be harnessed. These typically involve the incorporation of bipolar gradients and pulse sequences that modulate the internal gradient spectral frequencies in a way that removes the overlap with the desired diffusion encodings (Shemesh and Cohen 2011, Shemesh, Rosenberg et al. 2017). It is important to note that while these sequences mitigate the amplitude modulation of DDE, they do not assist in overcoming imaging artifacts due to strong internal gradients; these can be addressed by coupling the DDE diffusion weighting with many different image acquisitions strategies such as spatial encoding MRI, which reduce gradient-driven distortions.

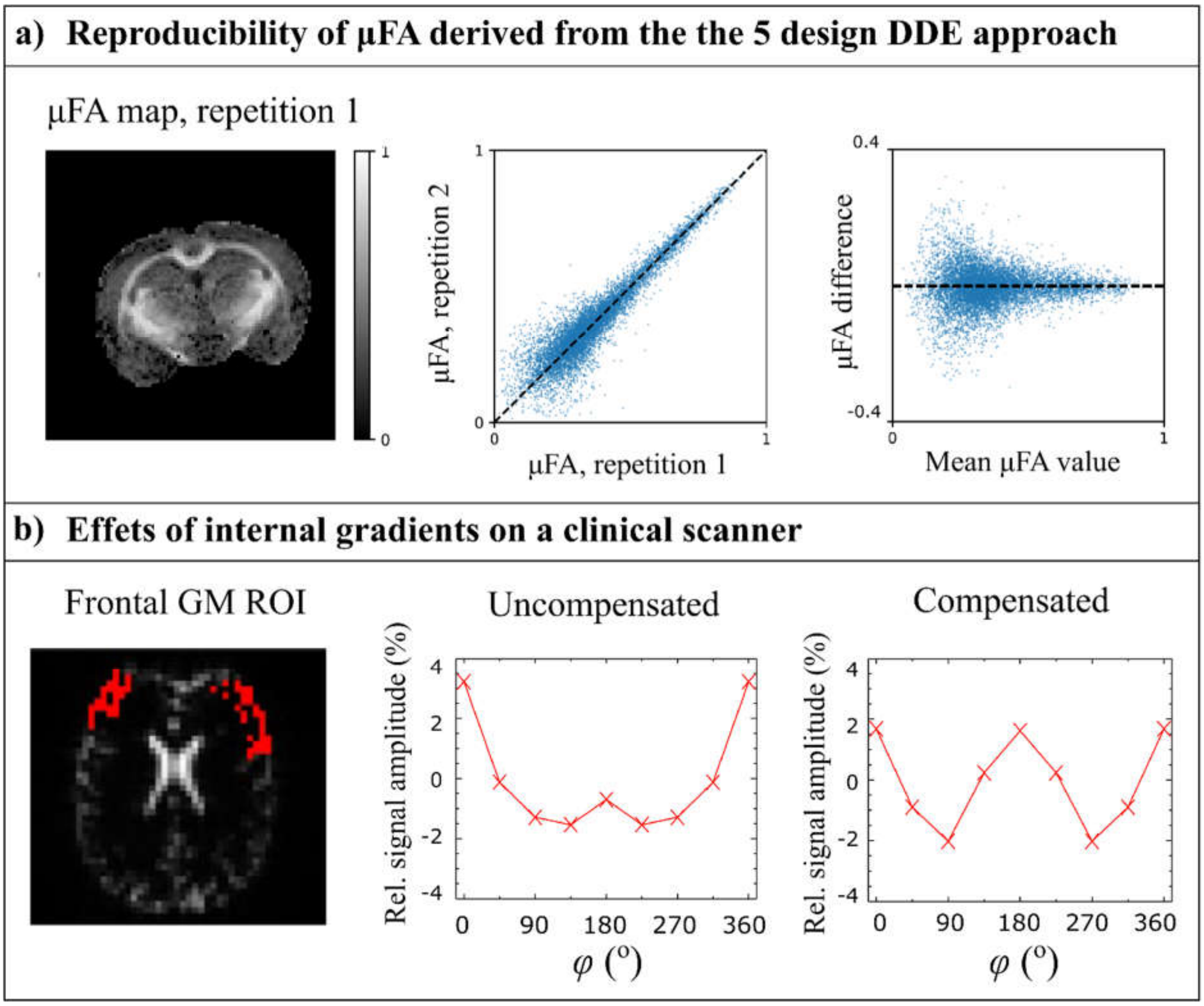

*Figure 10 a) Reproducibility analysis of µFA in the mouse brain, ex-vivo. The µFA values were derived from DDE measurements following the 5-design gradient scheme (left). The results show a very good correlation (r = 0.93) between the test-retest µFA values (middle), with larger differences occurring for voxels with lower microscopic anisotropy (right). Figure adapted from (Kerkelä, Henriques et al. 2020) b) Effects of internal gradients on a clinical scanner for cortical GM voxels (left). The signal amplitude modulation for compensated gradients shows the well-known cosine pattern (left), while the amplitude*

*modulation for uncompensated gradients is closer to a u-shaped curve (middle). Figure adapted from (Lawrenz and Finsterbusch 2019).*

**Effects from the MR-sequence**

Imaging gradients (in particular in STEAM settings, (Lasic, Lundell et al. 2018)), can induce significant and sometimes severe coupling (cross terms) with the DDE gradients. In particular, the slice selective pulses as well as crushers during the sequence can couple with the DDE gradients to produce unwanted modulations. Many strategies can be used to mitigate such effects, including applying opposite-sense gradients and combining the pairs ((Neeman, Freyer et al. 1990, Jara and Wehrli 1994)); applying nonselective DDE "preparation" prior to actual imaging (Komlosh, Lizak et al. 2008); and calculating effective b-matrices using all gradients applied during the sequence.

***Multiple effects – same sequence***

It should also be noted that a similar gradient configuration may also modulate qualitatively very different effects, e.g. the effects of exchange, restrictions and disperse flow all modulate experiments comparing parallel/antiparallel projections of the DDE encoding. While these different effects may occur at very different q-values or mixing times a certain simplified description of the encoding may not capture all flavours of the encoding. One example is the b-value, which may mirror diffusivity very differently depending on gradient duration or shape, separation etc. It is thus important to specify both b-values and diffusion times for any given experiment.

# 9 Summary and future directions

This review focused on the main applications of double diffusion encoding and described the theoretical background, validation experiments as well as the preclinical and clinical studies presented in the literature to estimate pore size and size distribution (Section 3), to decouple the effects of microscopic anisotropy from orientation distribution (Section 4), as well as to probe displacement correlation, study exchange effects and act as a filter to null fast diffusion components (Section 6). We also discussed the feasibility of using DDE acquisitions to estimate model-free rotational invariants as well as to provide orthogonal measurements for improving the estimation of biophysical model parameters (Section 5). Besides their application to MRI, DDE methodologies have also been used in metabolite MRS studies to probe the diffusion properties of specific cell types in the brain tissue (Section 7). Throughout the manuscript, previous clinical and preclinical applications of the different DDE-based techniques were reviewed. In comparison to more conventional SDE-based techniques, DDE methodologies provide multiple advantages in the characterization of microstructural properties, although, the applicability of DDE as a routine tool in the clinics is still limited by its higher complexity of implementation, higher SNR demands and usually longer acquisition times (Section 8).

With the recent developments in MR technology, there is an increasing interest in advanced diffusion sequences, including DDE, both from scientists and clinicians. We see this research field going forward with developments along the entire imaging pipeline, from theoretical description, sequence optimisation, implementation and preclinical and clinical applications. Some of these topics are already under scrutiny, for instance:

• Exploring the DDE time dependence and effects of the cumulant expansion beyond $q^4$, both from a theoretical and experimental perspective (Ianus, Jespersen et al. 2018, Jespersen, Fieremans et al. 2019, Jespersen, Olesen et al. 2019);

• Theoretical description and experimental implementation of different gradient waveforms and “non ideal” DDE pulse sequence effects (Jespersen, Fieremans et al. 2019);

• Optimisation of DDE acquisitions (and beyond) for estimating different compartments shapes, or the parameters of different biophysical model (Coelho, Pozo et al. 2019);

• Extending the CTI model to separate the non-Gaussian kurtosis sources in specific directions and/or effects beyond $q^4$ (Henriques, Olesen et al. 2020).

Other topics are more general, and we believe will be the focus of future DDE studies:

• Comparing DDE and q-space trajectory encoding method and understanding the pros and cons of the two approaches both from a theoretical and practical perspective;

• Expanding the application of DDE methods to different animal models and validating the diffusion metrics with gold standard values derived from techniques such as histology (Skinner, Kurpad et al. 2017, Komlosh, Benjamini et al. 2018);

• Improving the clinical translation (Yang, Tian et al. 2018, Kerkelä, Henriques et al. 2020);

• Studying the sensitivity/specificity of DDE metrics in a variety of diseases, e.g. (Lampinen, Szczepankiewicz et al. 2017, Yang, Tian et al. 2018);

• Exploring multidimensional diffusion acquisitions and combinations with relaxation techniques, as well as different numerical methods for solving the inverse problems (Benjamini and Basser 2016, de Almeida Martins and Topgaard 2018, Slator, Hutter et al. 2019, Reymbaut, Mezzani et al. 2020).

Finally, it seems that the time is ripe for exploring DDE’s potential both in preclinical as well as clinical settings, to characterize tissue microstructure in neurodegenerative diseases, stroke, cancer, as well as in development, learning and aging.

These are some of the research directions to be explored in future DDE studies, and we hope that this review will not only provide a summary of current DDE applications, but also a reference for the future development, validation and full clinical translation of DDE-based techniques.

**Acknowledgements**

The authors would like to acknowledge the following sources of financial support. AI is supported by European Union’s Horizon 2020 research and innovation programme under the Marie Skłodowska-Curie grant agreement No 101003390 and Champalimaud Centre for the Unknown, Lisbon (Portugal); RNH and NS are supported by European Research Council (ERC) (agreement No. 679058); MP is supported by Engineering and Physical Sciences Research Council (EPSRC EP/N018702/1) and UKRI Future Leaders Fellowship MR/T020296/1; SJ is supported by Danish National Research Foundation (CFIN), and the Danish Ministry of Science, Innovation, and Education (MINDLab); HL is supported by H2020 European Research Council, Grant/Award Number: 804746; Danish Council for Independent Research, Grant/Award Number:4093-00280B;

**Declarations of interest: none.**